\documentclass[fleqn,usenatbib]{mnras}

\usepackage{newtxtext,newtxmath}

\usepackage[T1]{fontenc}
\usepackage{ae,aecompl}

\usepackage{graphicx}	
\usepackage[utf8]{inputenc}
\usepackage{amsmath}	
\usepackage{soul,ulem}
\usepackage[dvipsnames]{xcolor}

\title[GJ 436 HD simulations]{GJ 436b and the stellar wind interaction: simulations constraints using Ly$\alpha$ and H$\alpha$ transits}

\author[C. Villarreal D'Angelo et al. ]{
Carolina Villarreal D'Angelo$^{1,2}$,\thanks{E-mail: carolina.villarreal@unc.edu.ar}
Aline A. Vidotto$^{1}$,
Alejandro Esquivel$^{3}$,
\newauthor
Gopal Hazra$^{1}$,
Allison Youngblood$^{4}$
\\
$^{1}$School of Physics, Trinity College Dublin, College Green, Dublin 2, Ireland.\\
$^{2}$Observatorio Astron\'omico de C\'ordoba - Universidad Nacional de C\'ordoba. Laprida 854, X5000BGR. C\'ordoba, Argentina.\\
$^{3}$Instituto de Ciencias Nucleares, Universidad Nacional Aut\'onoma de M\'exico, Apartado Postal 70-543, 04510 Ciudad de M\'exico, Mexico.\\
$^{4}$Laboratory for Atmospheric and Space Physics, University of Colorado, Boulder, CO 80303, USA
}

\date{Accepted 2020 December 7. Received 2020 November 27; in original form 2020 April 9}

\pubyear{2020}

\begin{document}
\label{firstpage}
\pagerange{\pageref{firstpage}--\pageref{lastpage}}
\maketitle

\begin{abstract}
The GJ 436 planetary system is an extraordinary system. The Neptune-size planet that orbits the M3 dwarf revealed in the Ly$\alpha$ line an extended neutral 
hydrogen atmosphere. This material fills a comet-like tail that obscures the stellar disc for more than 10 hours after the planetary transit. Here, we carry out a 
series of 3D radiation hydrodynamic simulations to model the interaction of the stellar wind with the escaping planetary atmosphere. With these models, we seek 
to reproduce the $\sim56\%$ absorption  found in Ly$\alpha$ transits, simultaneously with the lack of absorption in H$\alpha$ transit. Varying the stellar wind 
strength and the EUV stellar luminosity, we search for a set of parameters that best fit the observational data. Based on Ly$\alpha$ observations, we found a 
stellar wind velocity at the position of the planet to be around [250-460] km s$^{-1}$ with a temperature of $[3-4]\times10^5$ K. The stellar and planetary mass 
loss rates are found to be $2\times 10^{-15}$ M$_\odot$ yr$^{-1}$ and $\sim[6-10]\times10^9$ g s$^{-1}$, respectively, for a stellar EUV luminosity of 
$[0.8-1.6]\times10^{27}$ erg s$^{-1}$.
For the parameters explored in our simulations, none of our models present any significant absorption in the H$\alpha$ line in agreement with the observations.
\end{abstract}




\section{Introduction}

Just like the planets in our solar system, exoplanets are immersed in the (magnetised) wind of their host stars. In particular, gas-giant planets that orbit very 
close to their hosts (orbital distances $a_p<0.5 $ au) experience a large stellar irradiation flux that causes the heating and expansion of the upper layers of their 
atmosphere. The atmospheric material then escapes in the form of a `planetary wind' \citep{Lammer2003}. 

The physical process involved in the interaction between the stellar and the induced planetary wind leaves an imprint that can be observed during  planetary 
transits and can be modelled by means of numerical simulations. Comparing the model result with the observations allows us to constrain the parameters of the 
planetary system that are not directly determined by observations \citep{Villarreal2014, Vidotto2017, Villarreal2018, Kislyakova2019}.

The  Lyman $\alpha$ line has become an important tool in this sense. This line have shown that in certain cases the amount of neutral hydrogen leaving the 
planet atmosphere can obscure a substantial part of the stellar flux during the planetary transit. One of the most extreme cases is that of the Neptune-mass 
exoplanet GJ 436b. For this exoplanet, an excess in absorption was detected first by \cite{Kulow2014} and later confirmed by \cite{Ehrenreich2015}. In this work, 
observations with the STIS spectrograph on board Hubble revealed that the absorption in Ly$\alpha$ at mid-transit reached $56.3 \pm 3.5 \%$, almost 80 times 
larger than the absorption caused by the planet in broadband optical transit ($0.69\%$). The Ly$\alpha$ absorption was observed to  start $\sim3$ hours before 
mid-transit, and to end more than 10 hours after mid-transit \citep{Bourrier2016, Lavie2017}. This asymmetry indicates that the neutral hydrogen escaping GJ 
436b takes the form of a comet-like tail.  

The set of Ly$\alpha$ observations gathered for this exoplanet covering the planetary transit spans more than five years  (2012, 2013, 2014 and 2016 
\citep{Kulow2014, Ehrenreich2015, Lavie2017}). In most of them, a maximum absorption of  $\sim 60\%$ in the blue wing of the line (defined between [-120,-40] 
km s$^{-1}$) occurring close to mid-transit, appears stable in the different epochs. On the contrary, the red wing defined between [30,110] km s$^{-1}$, show a 
time variable absorption after mid-transit revealing the presence of transients either from stellar or planetary origin.  

Numerical models have shown that the observational features found in Ly$\alpha$ can be explained with an evaporating atmosphere that expands several 
planetary radii behind the planet \citep{Bourrier2016, Khodachenko2019, Kislyakova2019}. All these models assume that, to some extent, radiation pressure, 
charge exchange and the interaction with the stellar wind give rise and shape the neutral material that leaves the planet, giving the asymmetric shape of such 
absorption. Although it is clear by now that all of these physical process are in play, there is no consensus on which process play a major role (see 
\citealt{Bourrier2016} and \citealt{Khodachenko2019}). Despite the differences, the main features of the observed Ly$\alpha$ absorption profile are well 
reproduced by the different models, mainly the sharp ingress and a long egress observed in the blue wing. However, none of the models developed so far can 
simultaneously render all the data acquired for this exoplanet with a single set of parameters. Moreover, parameters like stellar wind velocity or stellar and 
planetary mass loss rate used to match the observational Ly$\alpha$ absorption of GJ 436b in the different models are still in disagreement, as we discuss next. 

Multi-species hydrodynamic models either in 1D \citep{Salz2016, Loyd2017} or 3D \citep{Shaikhislamov2018,  Khodachenko2019} agree with a planetary mass-loss rate of 
$3\times10^{9}$ g s$^{-1}$ and a planetary wind terminal velocity of 10 km s$^{-1}$ to reproduce the observations. On the other hand, the particle simulations of 
\cite{Bourrier2016} and \cite{Lavie2017} predict a planetary wind velocity between [50-60] km s$^{-1}$ and a mass-loss rate an order of magnitude lower. 

The stellar wind parameters, such as density and velocity at the interaction region, are by-products of the Ly$\alpha$ escape modelling. \citet{Bourrier2016} and 
\citet{Vidotto2017} estimated a stellar wind mass-loss rate between $[0.5-2.5]\times10^{-15}M_\odot$ yr$^{-1}$. \citet{Vidotto2017} found that an isothermal 
stellar wind temperature of [0.36-0.43] MK is needed to reproduce the local stellar wind velocity of 86 km s$^{-1}$, estimated by \cite{Bourrier2016}. Recently, 
\citet{Mesquita2020} presented  stellar wind models for GJ 436 that account for the presence of Alfv\'en waves to heat and accelerate the stellar wind. These 
authors derive a mass loss rate of $<7.6\times10^{-15}M_\odot$ yr$^{-1}$, consistent with \citet{Bourrier2016}, but a significantly higher stellar wind velocity with 
an upper limit of $800$ km s$^{-1}$ at the position of the planet. 

With the exception of \citet{Mesquita2020}, who modelled only the wind of GJ 436, all the estimated parameters for the GJ 436 system have used only the Ly$
\alpha$ observations to constrain the model results, as in-transit absorption in other spectral lines have not been detected. \cite{Cauley2017} searched for H$
\alpha$ in-transit absorption without success, despite the extraordinary absorption of the neutral hydrogen in the Ly$\alpha$ line. For the more heavy species like 
C$\mathrm{II}$ and Si$\mathrm{III}$, \cite{Loyd2017} were able to put an upper limit to the expected absorption based on a numerical model.

In this work we seek to constrain the stellar and planetary wind properties of GJ 436 by modelling the stellar and planetary wind interaction and computing 
synthetic Ly$\alpha$ and H$\alpha$ transits, to be compared to the observations. Table \ref{tab:1} summarises the physical properties of this system.
We employ a 3D radiative hydrodynamic code to model the propagation and interaction of stellar and planetary winds exploring different parameters. The stellar 
wind of the GJ 436 is determined by setting the coronal temperature of the star and the mass loss rate. The planetary wind is fully determined by the amount of 
EUV flux (F$_\mathrm{EUV}$), derived from observations, that the atmosphere receives. The wind from the planet is modelled with a 1D code that solves the 
atmospheric escape process from the lower planetary atmosphere \citep{Allan2019}, and provides the boundary conditions in the 3D model. The same 
F$_\mathrm{EUV}$ value is used in both the 1D and 3D models. 
In this way, a consistent set of initial parameters is used to explore the wind-wind interaction in this exoplanetary system. With this, we obtain the observational 
signatures in both H$\alpha$ and Ly$\alpha$ lines during transit, and compare those to previous observations.

\begin{table}
	\centering
	\caption{GJ 436 system parameters. Stellar parameters taken from \protect\citet{Torres2007} and \protect\citet{France2016}. Planet parameters taken from 
\protect\citet{vonBraun2012}}
	\label{tab:1}
	\begin{tabular}{llll} 
		\hline
		\hline
		 Object & Symbol & Value \\
		\hline
		\bf GJ 436&&\\
		Mass [$\mathrm{M}_\odot$]   & M$_\star$ &0.50\\
		Radius [$\mathrm{R}_\odot$] & R$_\star$ &0.45\\
        distance [pc] & d & 10.2\\
		\hline
		\bf GJ 436b &&\\
		Mass [$\mathrm{M}_\mathrm{J}$]   & M$_\mathrm{p}$ & 0.078\\
		Radius [$\mathrm{R}_\mathrm{J}$] & R$_\mathrm{p}$ & 0.36\\
		semi-major axis [au]             & a$_\mathrm{p}$ & 0.03\\
		Orbital period [d]               & $\tau_\mathrm{p}$ & 2.64\\
		Inclination [$^\circ$] & i & 86.6\\
		\hline
		\hline
	\end{tabular}
\end{table}

The paper is organised as follows: Section \ref{sim} introduces the numerical code with all the physical process involved. In this section we explain the boundary 
conditions used in the simulations to reproduce the stellar and planetary wind. Section \ref{res} presents the results from our models and Section \ref{espectro} 
the computation of the synthetic H$\alpha$ and Ly$\alpha$ lines. We present a summary of our results in Section \ref{summ} and a discussion about the caveats in our modelling approach -- for example, our model includes radiation pressure (investigated in Appendix \ref{appendix}), but does not include charge-exchange processes. We present our conclusions in Section \ref{conc}.

\section{Simulations} \label{sim}
To simulate the GJ 436 planetary system we employ the hydrodynamics/magnetohydrodynamics-radiative code {\sc guacho}\footnote{Freely available in https://
github.com/esquivas/guacho}. {\sc guacho} has already been employed in several works to simulate the interaction of stellar and planetary wind of the HD 
209458 system \citep{Esquivel2019,Villarreal2018,Schneiter2016}. We have made use of the hydrodynamic version of the code  that solves the following set of 
equations:

\begin{equation}
\frac{\partial \rho}{\partial t} + \nabla \cdot (\rho {\bf u})=0,
\label{eq:cont}
\end{equation}

\begin{equation}
\frac{\partial (\rho {\bf u})}{\partial t} + \nabla \cdot (\rho {\bf
  uu}+{\bf I}P)= \rho ({\bf g_p}+{\bf g_{e,\star}}) ,
\label{eg:mom}
\end{equation}

\begin{equation}
\begin{split}
\frac{\partial E}{\partial t} + \nabla \cdot [{\bf
    u}(E+P)]=G_\mathrm{rad}-L_\mathrm{rad} + 
    \rho \left( {\bf g_p}+ {\bf g_{e,\star}}\right) \cdot
     {\bf u},
\label{eq:ener}
\end{split}
\end{equation}
where $\rho$, {\bf u}, P, and E are the mass density, velocity,
thermal pressure and energy density, respectively. {\bf I} 
is the identity matrix, while $G_\mathrm{rad}$ and $L_\mathrm{rad}$ the
gains and losses due to radiation.
 ${\bf g_{e,\star}}$ and ${\bf g_p}$  are the  (effective) stellar and planetary gravitational acceleration. 
The total energy density and thermal pressure are related by an ideal gas equation of state $E = \rho \vert{\bf u}\vert^2/2 + P/(\gamma-1)$,
where $\gamma=5/3$ is the ratio between specific heat capacities.

The hydrodynamics equations (left hand side of Equations
\ref{eq:cont}--\ref{eq:ener}), are advanced with a second order
Godunov method with an approximate Riemann solver (HLLC,
\citealt{torobook}), and a linear reconstruction of the primitive
variables using the $\mathrm{minmod}$ slope limiter to ensure stability.

The numerical setup have the star and the planet within the computational domain with the planet orbiting around the star in the xz-plane. 
Our physical domain covers $[0.15,0.03,0.75]$ au ($\sim 65 \mathrm{R}_{\star}$ in the x-direction) divided in 880 $\times$ 176 $\times$ 460 cells in the x, y and 
z direction, respectively, as we only simulate half of the orbital plane.  With this, our resolution is about $1.7\times 10^{-4}$ au ($\sim0.07 \mathrm{R}_\star$). By 
simulating only half of the orbit, and imposing the axisymmetric nature in the numerical setup, we can save computational resources. Our simulations take 60 
core-hours in 84 cores to complete.

\subsection{Source terms}
The right hand-side of Equations (\ref{eq:cont}) to (\ref{eq:ener}) are our source terms, which represent all the external quantities that make the equations non-
conservative. In our numerical scheme, they are included after each timestep. These terms are described next.

\subsubsection*{Gravity $\&$ Radiation pressure}\label{grp}
In our simulations the gravity forces from the star and the planet are included as that of two point sources. We also include the effect of stellar radiation pressure 
as in \cite{Esquivel2019}, and so the gravity of the star is reduced by a factor proportional to the flux in the Ly$\alpha$ line and the neutral fraction. The effective 
stellar gravity is then:
\begin{equation}
   {\bf g_{e,\star}}=(1-\beta(v)\chi_n){\bf g_\star},
   \label{eq:grav_eff}
\end{equation}
with $\chi_n$ the total neutral fraction inside each grid cell and $\beta(v)$ a velocity dependent parameter proportional to the flux in the line. To compute $
\beta(v)$ we employ the formula presented in \cite{Lagrange1998} and the reconstructed flux of GJ 436 in the Ly$\alpha$ line taken from the MUSCLES survey 
\citep{France2016, Youngblood2016, Loyd2016}. The implementation of the radiation pressure in our simulations neglects self-shielding. Because of this,  the radiation pressure force can be overestimated. We analyse the consequence of this in Section \ref{discussion_radp_cexch} and address this point in more detail in  Appendix \ref{appendixA2}. 

\begin{figure}
    \centering
    \includegraphics[width=\columnwidth]{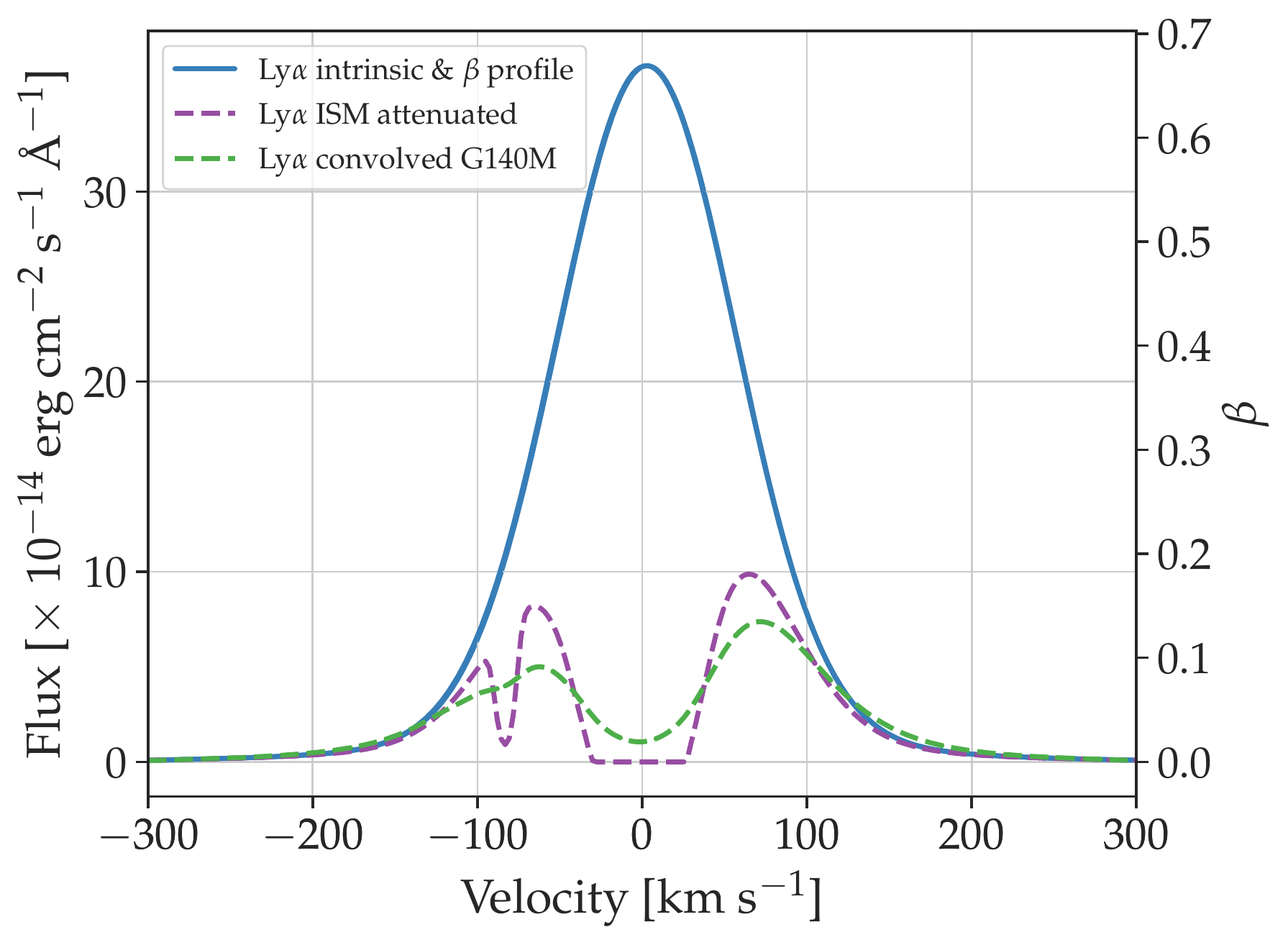}
    \caption{GJ 436 intrinsic Ly$\alpha$ flux at Earth obtained from the MUSCLES survey and the corresponding $\beta$ profile used to calculate the radiation 
pressure force in our simulations (blue solid line). Dashed lines correspond to flux attenuated by the ISM (violet) and then convolved with the G140M grating 
response of the HST STIS instrument (green).}
    \label{fig:lya_line}
\end{figure}

The intrinsic Ly$\alpha$ flux at Earth and the resulting $\beta$ profile used in our simulations is shown in Figure \ref{fig:lya_line} (solid blue line). The figure also 
shows the attenuated profile due to the ISM absorption (violet line) and the one resulting from the convolution with the G140M grating from HST (green line), 
that will be used to simulate an observation with the STIS instrument during transit with our models (Fig. \ref{fig:lya_models}). 

The Ly$\alpha$ intrinsic profile taken from \cite{Youngblood2016} uses an H$\mathrm{I}$ column density of  $\mathrm{N}_\mathrm{H\mathrm{I}}
=1.1\times10^{18}$ cm$^{-2}$, a velocity centroid v$_\mathrm{H_\mathrm{I}}= -4.1$ km s$^{-1}$ and Doppler parameter b$_\mathrm{H\mathrm{I}}= 8.6$ km 
s$^{-1}$ to model the ISM H$\mathrm{I}$ absorption. For the deuterium absorption, we use the ratio  D$\mathrm{I}/\mathrm{H}\mathrm{I}=1.5\times10^{-5}$ (as 
in \citealt{Wood2004}), the same velocity centroid as v$_\mathrm{H\mathrm{I}}$, and we assume b$_\mathrm{D\mathrm{I}}=$ b$_\mathrm{H\mathrm{I}}/
\sqrt{2}$. 

This reconstructed Ly$\alpha$ line has a total flux of F$_{\mathrm{Ly}\alpha} = (2.1 \pm 0.3) \times 10^{-13}$ erg cm$^{-2}$ s$^{-1}$, which gives a 
F$_{\mathrm{Ly}\alpha} = 0.92$ erg cm$^{-2}$ s$^{-1}$ at 1 au in agreement with \citet{Ehrenreich2015}. The luminosity values used in our simulations are 
based in this F$_{\mathrm{Ly}\alpha}$ value (see section \ref{boundarycond}).

{}

\subsubsection*{Radiative processes}
Within our models, the neutral hydrogen can be ionised by collisions or EUV photons and is allowed to recombine from an ionised state. To follow the change of 
neutral hydrogen within our simulations we solve the following equation after every hydrodynamic step: 

\begin{equation}
 \begin{split}
  \frac{\partial n_\mathrm{HI}}{\partial t} + \nabla.(n_\mathrm{HI}
  {\bf{u}})=  & (n_\mathrm{H}-n_\mathrm{HI})^2\alpha(T)\\
& -(n_\mathrm{H}-n_\mathrm{HI})n_\mathrm{HI}c(T)-n_\mathrm{HI}\phi,
 \end{split}
\label{eq:hrate}
\end{equation}

\noindent where $n_\mathrm{H}=\rho/m_\mathrm{H}$ is the total hydrogen number density ($m_\mathrm{H}=1.66\times10^{-24}\,\mathrm{g}$ being the 
hydrogen mass) and  $n_\mathrm{HI}$ is the number density of neutral hydrogen. $\alpha(T)$, c(T) are the `case B' recombination and collisional ionisation 
rates presented in Table \ref{tab:rates}. 
The photoionsation rate, $\phi$ is computed through the rate of ionising photons (S) that reach a cell volume. Hence, $\phi=\mathrm{S}/(n_\mathrm{HI}dV)$, 
where S$=\mathrm{S}_0\exp(-n_\mathrm{HI}a_0\mathrm{dl})$ is the ionising photon-rate which is attenuated as the photons are absorbed in their propagation 
through the grid, $a_0=6.3\times10^{-18}$ cm$^{2}$ is the photoionisation cross-section at the threshold frequency (13.6 eV/h) and dl is the path that the photon travels, 
taken in steps of one half of the cell size. S$_0$ is the initial photon-rate value set by the stellar luminosity in the EUV. In our models, we calculate the initial 
value of S$_0$ by dividing the total luminosity value in the EUV range by 13.6 eV. 

To launch the stellar photons, we employ the ray-tracing method described in \cite{Esquivel2013}, with $10^{7}$ photon-packages being ejected from the star at 
a random position from its surface, and in a random outward direction. Each photon package is followed and absorbed within a cell every time they find neutral 
material. As the photons moves, we keep track of the ionising rate at each cell and the energy deposited to the flow. Since we are simulating half of the planet orbit, 
hence half of the star, the value of S$_0$ is divided by 2 (with  all the photons imposed only from the hemisphere within the domain).

\begin{table}
    \centering
    \caption{Rate coefficients of `case B' recombination and collisional ionisation reaction implemented  our simulations.}
    \begin{tabular}{lll}
     \hline
        $\alpha(T)$ & $2.55\times10^{-13}(10^4/T)^{0.79}$ & \cite{Osterbrock1989} \\
        $c(T)$ & $5.83\times10^{-11}\sqrt{T}\exp(-157828/T)$&\cite{Osterbrock1989} \\
    
     \hline
    \end{tabular}
    \label{tab:rates}
\end{table}

\subsubsection*{Heating and Cooling}
The incoming stellar radiation will heat the planetary wind due to photoionisation. Assuming that all the photoelectrons produced during this processes will 
heat the gas, the volumetric heating rate can be expressed as 
\begin{equation}
 G_\mathrm{rad}= n_\mathrm{HI} \int_{\nu_0}^{\infty} \frac{4\pi J_\nu}{h\nu} a_\nu e^{-\tau_\nu}  h(\nu-\nu_0) d\nu,
  \label{heating1}
 \end{equation}
 where $n_\mathrm{HI}$ is the neutral density, $4\pi J_\nu$ is the average intensity of the stellar radiation,  $\nu_0$ is the threshold frequency (13.6 eV/h), $a_\nu$ is the 
photoionisation cross section and $\tau_\nu=\int a_\nu n_\mathrm{HI}dl$ is the absorption of the stellar flux as it moves within the neutral material. 
Equation (\ref{heating1}) can be solved numerically when the stellar spectral energy distribution is known. In our case, to reduce computational time, we treat the stellar flux as monochromatic, with this, the absorption cross section ($a_\nu = 6.3\times10^{-18} (\nu/\nu_0)^3)$ becomes $a_0 = 6.3\times10^{-18}$ cm$^{2}$  \citep{Osterbrock1989}. In this case,  Equation (\ref{heating1}) becomes
 \begin{equation} 
 G_\mathrm{rad}=n_\mathrm{HI} \phi  \epsilon,
 \label{heating}
\end{equation}
where $ \phi $ is the photoionisation rate and $\epsilon= (h\nu - h\nu_0)$ is the average energy gain per photoionisation. In our models, we assume $\epsilon \sim2$ eV, meaning that the stellar flux is composed of photons with energy of $\sim 15.6$ eV given a heating efficiency of $(h\nu - h\nu_0)/h\nu \sim 0.1$. Heating efficiencies for planets with hydrogen dominated atmospheres where calculated in the work of \cite{Shematovich2014}, where it was found that a value of 0.2 was almost constant within the planet thermosphere. On the other hand, \cite{Yelle2004} found that the heating efficiency 
decreased from 0.5 up to 0.1 within the planetary atmosphere, expecting to be small above $\sim 2 R_p$ due to the escape of photoelectrons from the planetary 
atmosphere. Our choice of a heating efficiency of $\sim 0.1$ reflects this idea since our boundary condition to launch the planetary wind start at 5 R$_p$.  

The assumption of a monochromatic flux is quite useful in 3D simulations, as it reduces computational time. This assumption has been used in a series of works (e.g.,  \citealt{MurrayClay2009, Debrecht2019, McCann2019}). For example, \cite{Debrecht2019} 
assumes that stellar photons have a single energy of 16 eV, which is  representative of the integrated EUV flux from the Sun \citep{Trammell2011}. Recently, \cite{Hazra2020} showed that a monochromatic flux gives a similar solution for the planetary wind as that computed using a non-monochromatic case, and so, our choice of a stellar 
monochromatic flux is fairly justified.

The non-equilibrium cooling function due to radiative processes, $L_\mathrm{rad}$, used in our models is taken from the work of \cite{Biro1995}:
\begin{equation}
    L_\mathrm{rad}=L_\mathrm{ion} + L_{\mathrm{Ly}_{\alpha}} +(1-f) [L_\mathrm{rec} + A(L_{\mathrm{O_\mathrm{I}}} + L_\mathrm{O_\mathrm{II}})] + f 
L_\mathrm{equil}.
    \label{cooling} 
\end{equation}
The total volumetric cooling rate is then the contribution from the cooling due to collisional ionisation ($L_\mathrm{ion}$), collisional excitation of the Lyman $
\alpha$ line ($L_{\mathrm{Ly}_{\alpha}}$) and recombination ($L_\mathrm{rec}$). 

Additionally, cooling due to the collisional excitation of the forbidden O$_\mathrm{I}$ and O$_\mathrm{II}$ lines ($L_{\mathrm{O_\mathrm{I}}}$, 
$L_\mathrm{O_\mathrm{II}}$) is included, and is multiplied by a factor $A=7.033$, to account for the contribution of other important ions that produce cooling such as C, N and S (see the work of \citealt{Biro1995} for details). We do not include photoionisation heating by these metals as our 3D simulations start at 5 R$_\mathrm{p}$. Photoionisation of metals could become important lower in the 
planetary thermosphere, but for a solar abundance composition the error introduced by neglecting them should be smaller than the uncertainty in the EUV flux value \citep{Salz2015, Salz2016}.

For temperatures higher than $5\times10^4$ K, the cooling function in Equation (\ref{cooling}) approaches the cooling of an ionised gas in coronal equilibrium 
($L_\mathrm{equil}$). This is done through the function $f(T)=\frac{1}{2}\left(\frac{1+\mathrm{tanh}(T-5\times10^4)}{500 \mathrm{K}}\right)$ which allows to 
switch from the low temperature regime to a higher one. The parametrisation of $L_\mathrm{equil}$ can be found in \cite{Raga1989} and includes the free-free 
cooling. 

Our cooling function is not valid for temperatures lower than $1\times10^4$ K and therefore, we turn off the cooling below this temperature. This may not be 
realistic as the planetary wind temperature at the boundary of the planet is in all models of the order of $10^3$ K. However, we assume that the contribution of 
the radiative cooling at this temperature is not significant and the material will cool mostly as a consequence of the planetary wind expansion (adiabatic cooling).
Note that, we do not include cooling by H$_3^+$, as molecular hydrogen is found in the lower layers of the atmosphere (<1$\mu$bar, \cite{Koskinen2013}), which are not included in our computation.

\subsection{Boundary conditions}\label{boundarycond}
The stellar and the planetary winds are treated as inner boundary conditions in our global 3D simulations. At the position where the wind is launched we set the 
values for temperature, velocity, neutral fraction and total density (through the corresponding mass-loss rate) for a purely hydrogen fluid. These values are then 
re-imposed at every time step. The outer boundary conditions, i.e. the boundaries at the end of the mesh, are treated as outflow except for the boundary at the x-
axis. This boundary is set to a reflective boundary to mimic the axisymmetric nature of the system, as we only model half of the orbital plane.

\subsubsection*{Stellar wind and photoionising flux}\label{stellarwind}
The boundary conditions for the stellar wind are given by a Parker wind solution for a coronal temperature $\mathrm{T}_\star$: 
\begin{equation}
   v_r  \exp\left[-0.5\left(\frac{v_r}{c_s}\right)^2\right]= c_s \left(\frac{r_\mathrm{c}}{r}\right)^2 \exp\left(\frac{-2 \mathrm{r}_c}{r} + \frac{3}{2}\right).
 \label{eq:parker}
\end{equation}
Here $r_\mathrm{c}= \mathrm{GM}_\star / (2c_s^2)$ is the position of the critical point where the velocity of the wind equals the sound speed 
$c_s=\sqrt{\mathrm{R}_g \mathrm{T}_\star/\mu}$, with $\mathrm{R}_g$ the gas constant and $\mu=0.5$ the mean atomic weight of the particles in units of 
$m_\mathrm{H}$.
In this work, we vary the coronal temperature of the star between $1$ and $3$ MK to simulate different stellar wind strengths at the planet position. In the Parker wind model, the higher the 
temperature of the wind, the higher is the stellar wind velocity. We chose temperatures in this interval to cover values previously suggested in the literature. For 
example, in \cite{Vidotto2017}, an isothermal temperature of the stellar wind is estimated to be $0.41$ MK, while in \citet{Mesquita2020}, the predicted stellar 
wind temperature at the planetary orbit is $\lesssim 1.7$ MK, but can reach higher values at larger distances from the star. 

In our models, the launching radius, $r_\mathrm{sw}$, for the fully ionised stellar wind is set above or at the sonic point and the velocity at this radius is determined by Equation (\ref{eq:parker}). The choice of the launching radius help us save computational resources and allow us to use an adiabatic index ($\gamma=5/3$) to properly model shocks. More sophisticated stellar wind models include extra terms in the energy and momentum equations to heat and accelerate the stellar wind from the stellar radius when adopting $\gamma=5/3$ \citep{Mesquita2020}. 

Note that the Parker solution is isothermal {\bf ($\gamma=1$)}, while our models assume an adiabatic index. This means that the stellar wind cools down as it 
expands in the grid, with the highest temperature at $r_\mathrm{sw}$, and thus the temperatures quoted at the orbital distance of the planet are lower than their 
injection values of $1$ to $3$ MK (see Table \ref{tab:2}). 

The density at the boundary is set by the adopted mass-loss rate of the star. Previous works (\citealt{Vidotto2017,Mesquita2020}) estimated for GJ 436 a mass-loss 
rate of [0.03 -- 0.8]  $\dot{\mathrm{M}}_\odot$, with the solar mass-loss rate given by $\dot{\mathrm{M}}_\odot=2\times10^{-14}$ M$_\odot$ yr$^{-1}$. Thus, we 
assume  $\dot{\mathrm{M}}_\star=0.1\dot{\mathrm{M}}_\odot$ in our models. For a given mass-loss rate and considering an isotropic wind, this implies that 
the density at the wind boundary is $\rho(r_\mathrm{sw}) =\dot{\mathrm{M}}_\star/(v(r_\mathrm{sw}) \, 4\pi r_\mathrm{sw}^2)$, with $v(r_\mathrm{sw})$ 
determined according to the chosen stellar wind temperature (Equation \ref{eq:parker}).

The photoionising flux is simulated with the emission of photons from the surface of the star. We launch a total of $10^7$ photons packages in random directions 
from random positions on the stellar surface. The initial photon-rate (S$_0$) is set through the EUV luminosity L$_\mathrm{EUV}$ adopted for GJ 436. \cite{Ehrenreich2015} 
estimated a luminosity around $[2.8-3]\times10^{27}$ erg s$^{-1}$ (calculated in the range [124 -- 912] \AA ). Based on the Ly$\alpha$ line and the scaling 
relations from \citet{Linsky2014}, \citet{Youngblood2016} estimated a L$_\mathrm{EUV}=1.7\times10^{27}$ erg s$^{-1}$ in the same wavelength range. More 
recently, \citet{Peacock2019} derived a luminosity value of $4.9\times10^{27}$ erg s$^{-1}$, calculated in the range [124 -- 912] \AA .
 
For our models, we have chosen to explore three different values of  L$_\mathrm{EUV}=[0.8,1.6,4]\times10^{27}$ erg s$^{-1}$, which span a factor of 5 in 
luminosity and cover the range of literature EUV estimates for this star. Based on these values, we have named our models with the letter L for low, M for 
moderate and H for high, respectively. For the three luminosity values we chose, the photoionisation time of hydrogen near the interaction region (shock position), in the star-planet direction, varies from 28 to 8 hours
for the lowest and the highest L$_\mathrm{EUV}$ values respectively. These values change to 8 and 2 hours in the planetary tail.
We note that we do not change the Ly$\alpha$ flux used to calculate the radiation pressure in our models while varying L$_\mathrm{EUV}$, because this value 
is constrained by direct observations to be relatively constant over a period of several years (e.g., \citealt{Ehrenreich2011, France2013, Youngblood2016, Lavie2017}).

All the physical values adopted for the star and its wind are presented in Table \ref{tab:2}, together with the resulting values of the stellar wind at the planet orbit. 

\begin{table}
    \centering
    \caption{Physical quantities adopted for the star and its wind. All the stellar wind models assume a mass-loss rate of $2\times 10^{-15} M_\odot$/yr. The 
columns are: the model name, the stellar wind temperature at its launching radius, the EUV luminosity of the star, the velocity, temperature and density of the 
stellar wind at the orbital radius of the planet. Model names indicates the intensity of the stellar EUV luminosities: H (high), M (moderate) and L (low), and the 
chosen stellar wind temperature: 1 correspond to T$_{\star}$ =1 MK and 3 to T$_{\star}$=3 MK. }
    \begin{tabular}{lccccccc}
         \hline
         Model& $T_{\star}$  & $L_\mathrm{EUV}$ &$v (a_p)$  &$T (a_p)$   & $n_\mathrm{H} (a_p)$ \\
     - & [$10^6$K] & $ [10^{27}$ erg/s] &[km/s]  &[$10^5$K] & [cm$^{-3}$]\\
         \hline 
         L1& 1 & 0.8 &254 & 4  & 11661\\
         M1& 1  & 1.6     &254 & 4  & 11661\\
         H1& 1  & 4   & 199 & 2  & 1542 \\
         L3 & 3  & 0.8  &465 & 3  & 634\\
         M3 & 3   & 1.6 &465 & 3  & 634\\
         H3 & 3   & 4   &465 & 3  & 634\\
         \hline
    \end{tabular}
    \label{tab:2}
\end{table}

\subsubsection*{Planetary wind}\label{1dmodel}
Similar to the stellar wind, the planetary wind is launched at a given radius, r$_\mathrm{pw}$, where the velocity, temperature, density and ionisation fraction are 
set. These parameters are taken from the 1D hydrodynamic atmospheric escape model presented in \cite{Allan2019} and rendered uniformly around the 
planet at the launching radius. In this model, a planetary wind is generated from the heating and expansion of the upper planetary atmosphere as a 
consequence of the photoionising EUV flux from the star. The model takes as inputs the mass of the star, mass and radius of the planet, as well as the stellar 
EUV flux. For the escape model we also assume a density and a temperature at the bottom of the planetary atmosphere, which do not have a strong 
impact on the planetary wind profiles (see \citealt{Allan2019} and reference therein). 

\begin{figure}
    \centering
    \includegraphics[width=0.7\columnwidth]{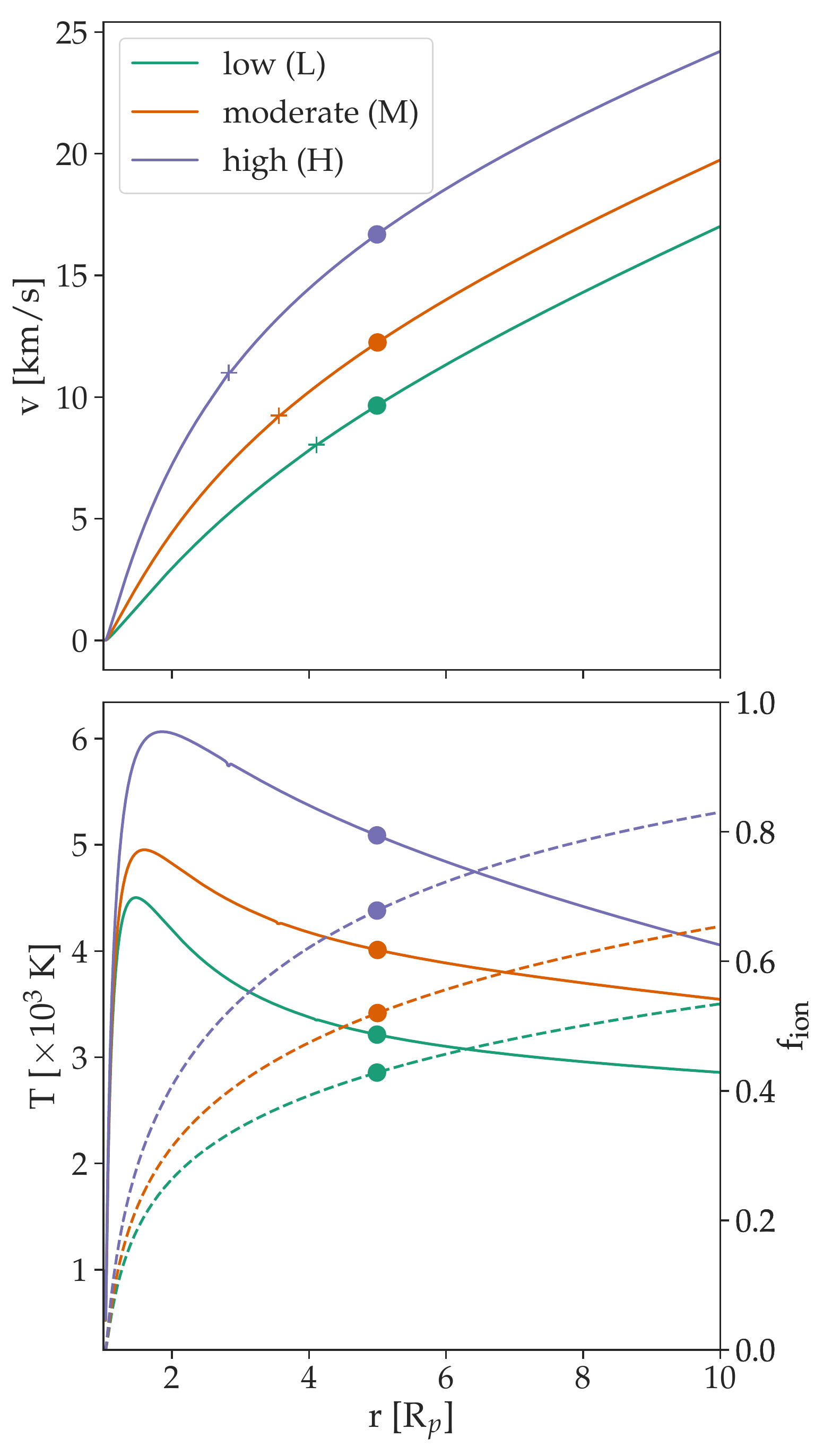}
    \caption{Planetary wind profiles for three different values of the stellar EUV flux (low, moderate and high) from the 1D atmospheric model of \citet{Allan2019}. 
Top: radial velocity. Bottom: Temperature and ionisation fraction. The dot represents the position of the boundary condition in the 3D model. The crosses in the 
velocity profile show the position of the sonic point.}
    \label{fig:pwprof}
\end{figure}

Using the values of L$_\mathrm{EUV}$ shown in Table \ref{tab:2}, we create 3 different planetary wind models with the 1D escape model. 
The radial profiles of velocity, temperature and ionisation fraction (f$_\mathrm{ion}$) for the 3 models are shown in Figure \ref{fig:pwprof}.  In the figure, 
model names correspond to the high (H), moderate (M) and low (L) stellar EUV luminosity. The values used at r$_\mathrm{pw}$ are marked with a dot and also 
stated in Table \ref{tab:models}. In the velocity profile the position of the sonic point for each of the planetary wind models is marked with a cross. 
Note that to 
keep the resolution in the 3D simulation in a reasonable value, we launch the planetary wind beyond the sonic point.
The planetary wind models generated in this way have an ionisation fraction of $[0.4,~0.5,~0.7]$ at  r$_\mathrm{pw}$ and a mass loss rate of $[5.5,~9.8,~20] 
\times10^9$ g s$^{-1}$ for the low, moderate and high L$_\mathrm{EUV}$, respectively. 

We can compare our results with other atmospheric escape models. For example, \cite{Salz2016}  estimated a mass-loss rate of $ 4 \times 4.4\times10^9 = 1.8 \times 10^{10}$ g s$^{-1}$ for a L$_\mathrm{EUV}=1.4\times 10^{27}$ erg s$^{-1}$ (\citeauthor{Salz2016}'s expression for $\dot{\mathrm{M}_p}$ has a factor of 1/4 compared to ours). Their model should be compared to our model M, as they have similar L$_\mathrm{EUV}$ values. When comparing to our value of $9.8\times 10^9$ g s$^{-1}$, we see that our mass-loss rate is comparable (within less than a factor of 2).
For similar L$_\mathrm{EUV}$ values, \citet{Loyd2017} and \citet{Shaikhislamov2018} found mass-loss rates of $3.1 \times10^9$ g s$^{-1}$ and $4 \times10^9$ g s$^{-1}$, respectively. Again, comparing to our model M, we see that their values are similar to ours (about a factor of 2 smaller).

\begin{table*}
	\centering
	\caption{Models boundary conditions. The model names correspond to the high (H), moderate (M) and low (L) stellar luminosity and to the high (3MK) and 
low (1MK) stellar wind base temperatures.}
	\label{tab:models}
	\begin{tabular}{lllllll}
		\hline
		\hline
		  Parameter                & L1 & L3 & M1 & M3 & H1 & H3 \\
		\hline
		{\it Stellar wind} &&&&&&\\
	    r$_\mathrm{sw}$ [R$_\star$]     & 7.9& 2.6& 7.9& 2.6 & 5.3 & 2.6\\
		T$_\star[10^6$ K]& $1$ & $3$ & $1$&  $3$ & $1$ & $3$\\
		$\dot{\mathrm{M}}_\star$ [$\dot{\mathrm{M}}_\odot$] & $0.1$ & $0.1$ & $0.1$ & $0.1$ & $0.1$ & $0.1$ \\
		v$_\mathrm{sw}$  [km/s]      & 181& 313& 181& 313&130& 313\\
		L$_\mathrm{EUV}$ [$10^{27}$ erg/s] & 0.8 & 0.8 & 1.6 & 1.6 & 4 & 4\\
		S$_0 [10^{36}$ s$^{-1}$] & $5.0$ & $5.0$ & $9.6$ & $9.6$ & $24.3$ & $24.3$\\
		\hline
		{\it Planetary wind} &&&&&&\\
		r$_\mathrm{pw}$ [R$_\mathrm{p}$]  & 5 & 5& 5 & 5& 5& 5\\
		T$_\mathrm{pw}$ [K] &  $3212$ &$3212$ & $4008$ & $4008$ & $5086$ & $5086$\\
		$\dot{\mathrm{M}}_\mathrm{p}[10^9$ g/s] & $5.5$ & $5.5$ & $9.8$ & $9.8$                      &$20$&$20$\\
		v$_\mathrm{pw}$ [km/s] & $9.7$ & $9.7$& $12.0$& $12.0$ & $16.7$& $16.7$\\
		f$_\mathrm{ion,pw}$ & $0.43$& $0.43$& $0.52$ & $0.52$& $0.68$& $0.68$\\
		\hline
		\hline
	\end{tabular}
\end{table*}

\section{3D modelling results}\label{res}
In order to characterise the environment around GJ 436b, we run six simulations. Three of them explore different values of L$_\mathrm{EUV}$ from the star, 
giving rise to three different planetary winds. These simulations are labelled with the letters L, M and H. For each of these simulations we adopt two different 
stellar wind models corresponding to two values of stellar  wind base temperature: 1 and 3 MK. The initial conditions used in our models are presented in Table 
\ref{tab:models}. 

\begin{figure*}
    \centering
    \includegraphics[width=\textwidth]{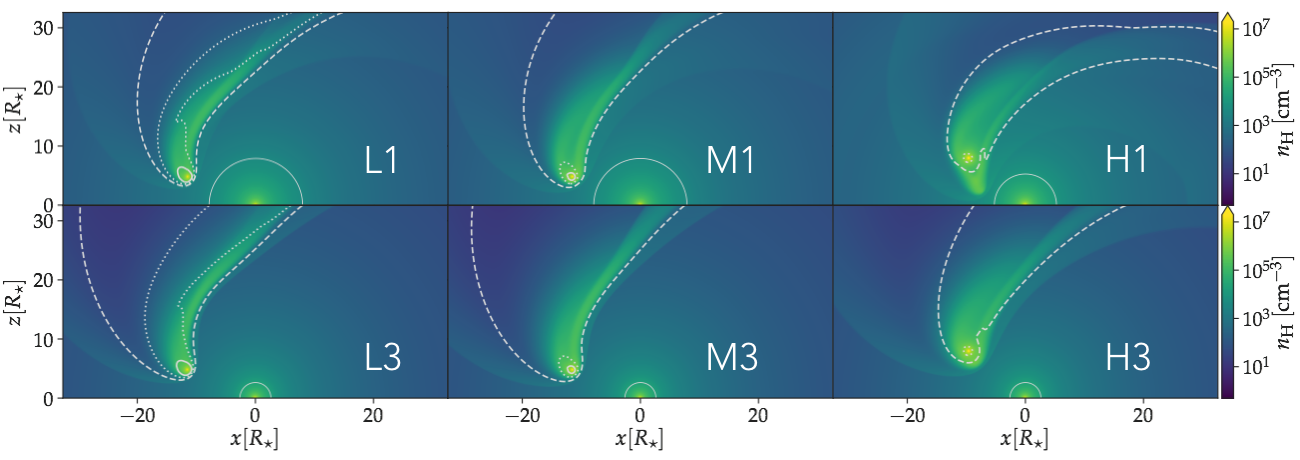}
   \caption{Cut in the orbital plane for t= 97200 s except for models H1 and H3 were t= 86400 s. The plots show the distribution of total hydrogen density for all 
the models. Top row: models with T$_\star=1$ MK. Bottom row: models with T$_\star=3$ MK. From left to right models with increasing stellar EUV flux. The 
white half-circle shows the launching radius of the stellar wind. The contours levels shown the ionisation fraction of 0.6, 0.8 and 0.99 from inside to outside. }
    \label{fig:neutralden}
\end{figure*}

Figure \ref{fig:neutralden} shows cuts of neutral hydrogen distribution in the orbital plane (xz) for all our simulations. The radius from where the stellar wind is launched, 
r$_\mathrm{sw}$, is marked with a white circle around the position of the star (origin of coordinates). The models are evolved until they reach a steady state and 
we show in Figure \ref{fig:neutralden} the evolution of the simulation up to $t=97200$ s, except for models H1/H3 where $t= 86400$ s for reasons we will present 
later.  This corresponds to a temporal evolution of approximately 2/5 of an orbit.

\begin{figure*}
    \centering
    \includegraphics[width=2\columnwidth]{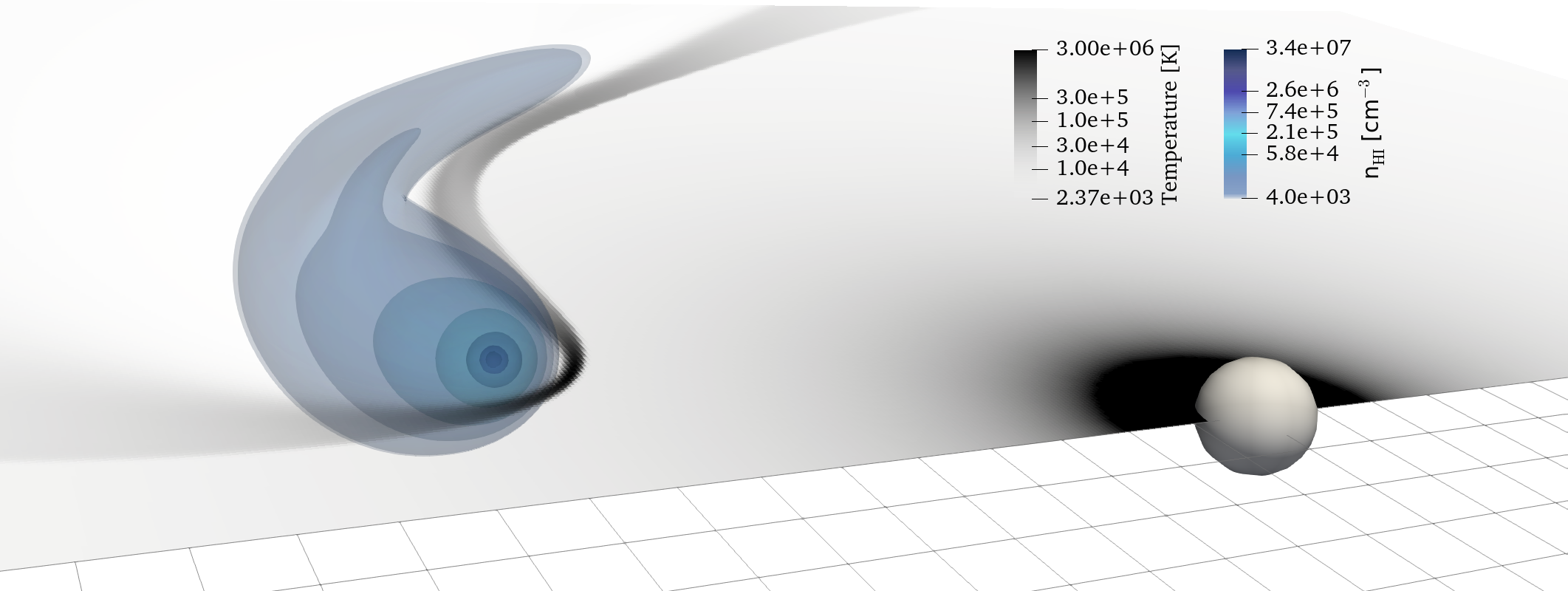}
    \caption{3D render view of model L3 at t= 97200 s. The orbital plane is coloured with temperature, while the contours around the planet represent the neutral H density. The shock is the high temperature region in front of the planet.}
    \label{fig:3D}
\end{figure*}

As the planetary and stellar winds expand and interact in our simulations, common features arise in all cases. On one hand, a shock is formed at the position where the two winds meet. In the shocked region, the temperature increases to around 1 and 3 MK (depending on the model), as is visible from Figure \ref{fig:3D} where temperature contour is shown in the orbital plane for model L3, together with iso-surfaces of neutral hydrogen around the planet. These high temperatures ionised the planetary material shaping the region of neutrals between the star and the planet. 

Another common feature is the development of a comet-like tail of escaping material trailing the planet. 
The extension of this tail vary according to the stellar wind strength. 
A stronger stellar wind (T$_\star=3$ MK, bottom row of Fig. \ref{fig:neutralden}) pushes the material in the tail towards the radial direction, whereas a slower 
stellar wind (T$_\star=1$ MK, top row of Fig. \ref{fig:neutralden}) allows the material to remain in the orbital path (i.e., the tail is more curved along the $\phi$ 
direction). 

The amount of neutral material in the tail is controlled by the stellar EUV flux. A higher L$_\mathrm{EUV}$ increases the value of the ionisation fraction at the 
boundary where the planetary wind is launched (see Table \ref{tab:models}). Then the neutral material that eventually escapes within the tail becomes ionised 
when interacting with the stellar photons. 
The overall result is that a higher stellar L$_\mathrm{EUV}$ produces a more ionised tail. This can be seen if we focus on the white contours on Figure 
\ref{fig:neutralden} that show the values of ionisation fraction $[0.6,~0.8,~0.99]$, represented by the solid, dotted and dashed lines around the planet, 
respectively.

As is visible from Figure \ref{fig:neutralden}, model H1 (which is the one with the weakest stellar wind and the strongest planetary wind) developed a stream of 
material from the planet in the direction of the star. This is the result of a stellar wind being incapable of halting planetary material that moves towards the star.
Hence, the planetary wind that gets more acceleration due to the high EUV flux is also accelerated by the stellar gravity and falls into the star ahead of the 
planet. This effect has already been studied by \citet{2017MNRAS.466.2458C} and \citet{Matsakos2015}.

This model has the higher planetary wind velocity and temperature at the launching radius (r$_\mathrm{pw}$) due to the high stellar L$_\mathrm{EUV}$ (see 
Fig. \ref{fig:pwprof}). 
The stellar wind in model H1 is imposed at a smaller radius (r$_\mathrm{sw}=\mathrm{r}_c$) with respect to models L1 and M1. We choose this radius to let the 
simulation evolve at roughly the same physical time of all the others models, before the planetary material reaches the stellar wind boundary. Because of this, 
the velocity and temperature of model H1 are lower at the planetary position than models with the same initial base temperature, and so, this model has a lower 
total stellar wind pressure than models L1 and M1 (see Table \ref{tab:2} and Section \ref{stellarwind}).

It is important to point out that the change in the position of the boundary condition for the stellar wind does not affect the outcome of the infalling material. We 
saw the same result for a boundary at 1.5 $r_\mathrm{c}$ (not shown here).

\section{Spectroscopic transits - synthetic line profiles}\label{espectro}
\subsection{Ly$\alpha$ calculation}\label{lyacalc}
To calculate the Ly$\alpha$ absorption produced by our models, we compute the optical depth as a function of velocity along the line of sight (LOS) for each cell 
in the grid accounting for the orbital inclination of GJ 436b (i=86.6$^\circ$):
\begin{equation}
 \tau_v= \int_{r_0}^{r}  \sigma_0 n_\mathrm{HI} W(a,v) dr_\mathrm{LOS}.
\end{equation}
Here, $\sigma_0$ is the absorption cross-section coefficient in the Ly$\alpha$ line at the threshold frequency (13.6 eV/h), n$_{\mathrm{HI}}$ is the neutral density, $W(a,v)$ is the Voigt line 
profile and $r_\mathrm{LOS}$ is the path along the LOS. The integration is done from above the stellar wind launching radius to the end of the computational 
mesh.

The Voigt profile $W(a,v)= H(a,v)/\sqrt\pi$ is written in terms of the Humileck function which depends on the damping parameter $a=A_{ij} \lambda_0/(4\pi v_{th})
$, and $v=(v_r-v_\mathrm{LOS})/v_{th}$. Here $A_{ij}$ is the transition rate, $\lambda_0=1215.67$ \AA is the central wavelength of the line, $v_{th}=\sqrt{2k_BT/
m_\mathrm{H}}$ is the thermal width, $v_r$ is the radial velocity which goes from $+300$ to $-300$ km s$^{-1}$ in 250 bins and $v_ \mathrm{LOS}$ is the velocity 
in the LOS direction. The line parameters are shown in Table \ref{tab:lines}. 

\begin{table}
    \centering
    \caption{Physical parameters adopted in the computation of the synthetic Ly$\alpha$ and H$\alpha$ transits.}
    \begin{tabular}{lll}
    \hline
      Parameter  & Ly$\alpha$ & H$\alpha$\\
    \hline
       $\lambda_0$ [cm]    & $1.215668\times 10^{-5}$& $6.56279\times 10^{-5}$\\
       $\sigma_0$ [cm$^2$] & 0.01105    & 0.017014  \\
       A$_{ij}$ [s$^{-1}$] & $6.27\times 10^{8}$     & $4.4101\times 10^{7}$  \\
    \hline
    \end{tabular}
    \label{tab:lines}
\end{table}

To calculate the transmission spectra we integrate the normalised intensity within the stellar radius, neglecting any limb-darkening variation and so, the 
absorption spectra is \\ $1- I_v/I_\star = 1- e^{-\tau_v}$. The total absorption is then the integration in the velocity range $\pm 300$ km s$^{-1}$.

Because we only simulate half of the orbit and our analysis is done at the end of our simulation, 
we emulate the temporal evolution of the transit by 
rotating the LOS direction (z-axis) around the $-y$-axis (out of the page) in a clockwise direction, i.e. towards the x-axis. In this way, every angle swept by the 
LOS direction represent a time that can then be measured from mid-transit (t=0). We stop our line profile calculations at an angle of 45$^\circ$, which corresponds to 
4 hours after mid-transit. After this, part of the comet-like tail would have been outside the simulation domain and thus not contributing to the absorption line.

\begin{figure*}
 \includegraphics[width=\columnwidth]{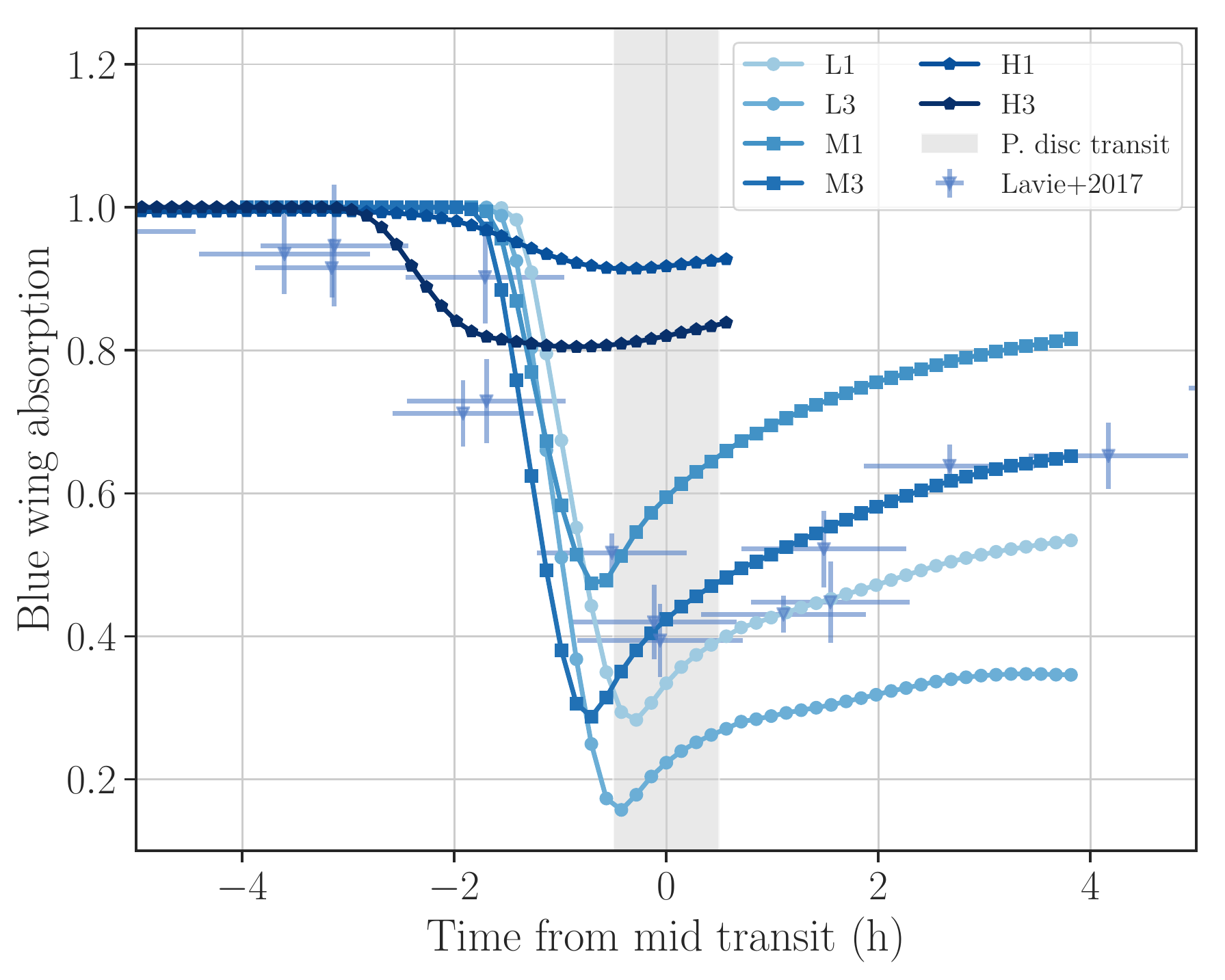}
  \includegraphics[width=\columnwidth]{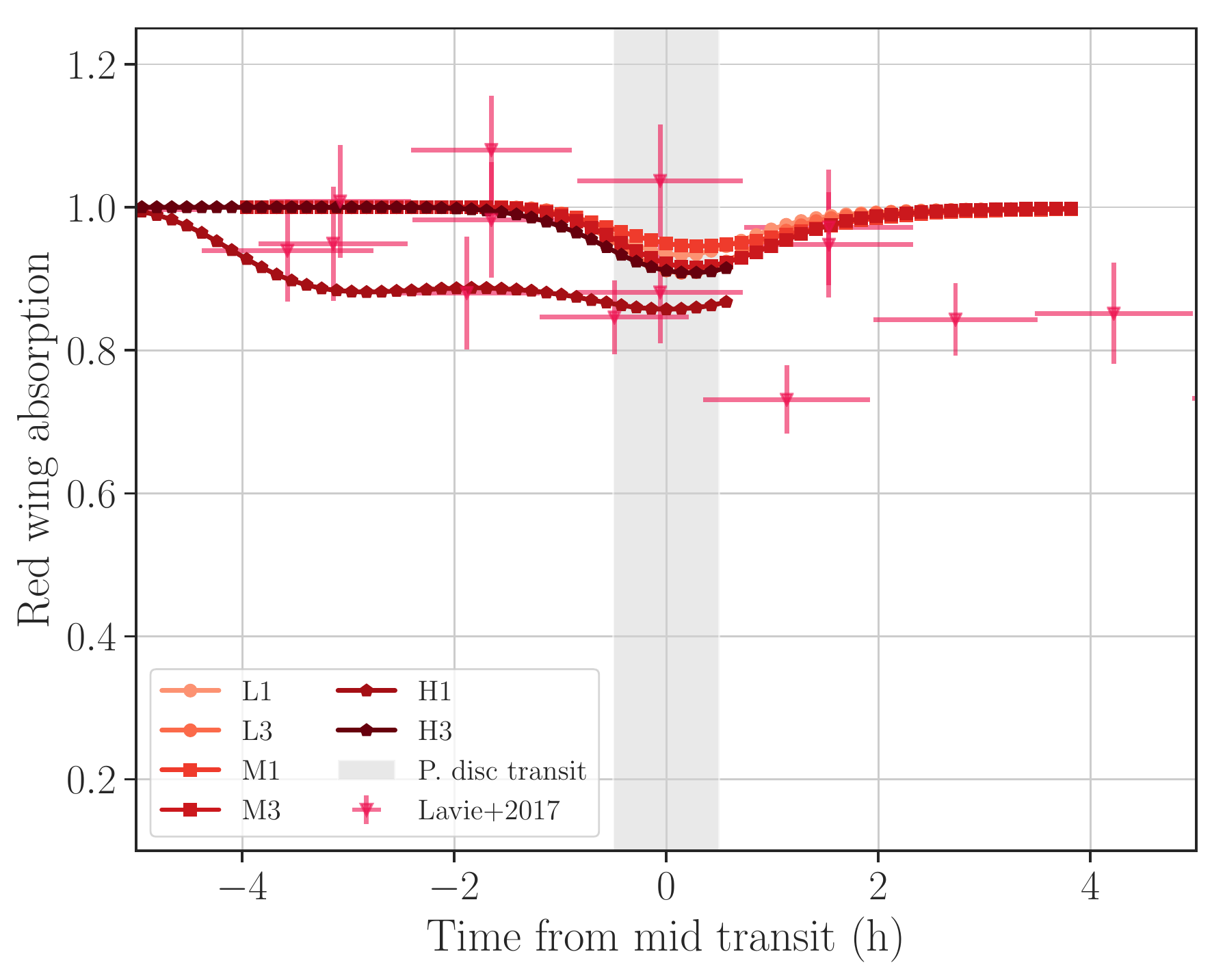}
    \caption{Blue and red wing Ly$\alpha$ absorption as a function of time for all the models together with the observations from \protect\cite{Lavie2017}. The 
grey band shows the duration of the optical transit.}
    \label{fig:transit}
\end{figure*}

Following the approach from \cite{Lavie2017}, we integrate the absorption in the Ly$\alpha$ line in two velocity ranges: the blue wing [-120,-40] km s$^{-1}$ and 
the red wing [30,110] km s$^{-1}$. The results of the two integrations are shown in Figure \ref{fig:transit} for all the models as a function of time from mid-transit. 
The figure also shows the duration of the optical transit in the grey stripe and the observations made in the different epochs extracted from \cite{Lavie2017} (blue 
and red triangles).

\subsubsection{Blue wing absorption ([-120,-40] km s$^{-1}$)}
As shown in the left panel of Figure \ref{fig:transit}, models with low and moderate L$_\mathrm{EUV}$ show the best agreement with the observations.  Models 
L1/L3 and M1/M3 present highly asymmetric lightcurve with a long egress phase, due to the amount of neutral material within the comet-like tail. The duration 
and maximum depth of the absorption produced by this material is in good agreement with the observational results.
These models, however, cannot reproduce the early ingress absorption observed 2 hours before mid-transit. This is because the neutral material does not 
extend ahead of the planet, implying that, in these simulations, the interaction between the stellar and planetary wind is produced close to the planet position. We 
also see that the slope in the lightcurve at the ingress seems to depend on the strength of the stellar wind, a higher coronal temperature (i.e. a stronger wind) produces a steeper slope. 

The maximum absorption values found for models L1/L3 and M1/M3 span a range from 50 to $80\%$. This spread is related to the amount of neutral material 
that escapes from the planet, which is larger for the lowest stellar EUV flux (L$_\mathrm{EUV}=0.8\times 10^{27}$ erg s$^{-1}$). 

Models with the highest stellar flux (L$_\mathrm{EUV}=1.6\times 10^{27}$ erg s$^{-1}$), H1 and H3, have the shallowest absorption, with values smaller than 
$20\%$. Although these models present an absorption that begins around 2 hours before mid-transit (being model H3 the earliest one), these models do not 
shown a long absorption duration, neither the $\sim 56\%$ of absorption seen in the observations at mid-transit. This is due to the high ionisation fraction of the 
planetary wind, caused at the same time by the high L$_\mathrm{EUV}$. Even though these models also show an extended comet-like tail, the amount of 
neutral material in this tail is too low to create sufficient absorption in this part of the line. 

In all cases, models with the same EUV flux have a largest absorption when the stellar wind temperature is highest. This is because a stronger stellar wind 
confines more efficiently the neutral material in the radial direction and then the column density of this material increases in this direction.  

\subsubsection{Red wing absorption ([30,110] km s$^{-1}$)} 
The right panel of Figure \ref{fig:transit} shows the absorption in the red wing. All of our models present a  more symmetric lightcurve with a maximum absorption 
around $10\%$ occurring in most of the cases half-hour after mid-transit. Only model H1, which is the one where planetary material falls towards the star, shows 
a deeper absorption that start about 4 hours before than the optical transit.
Contrary to the observations, none of our models reproduce the absorption detected between 2 and 4 hours after mid-transit, implying that the neutral material 
moving towards the star remains up to a few planetary radii around the planet.

The absorption produced by our models can, in some cases, reproduce the absorption observed in the different wings of the Ly$\alpha$ line. But, individual 
models  by themselves can not fit all the observational data in both wings simultaneously. Given that the late absorption found in the red wing is from a different 
epoch than the early ingress found in the blue wing, we speculate that some of these features, specially the early absorption in the blue wing and the late 
absorption in the red wing, could be due to time-dependent events in the system that influences the interaction with the planetary wind. 

\subsubsection{Synthetic STIS line observation} 
\begin{figure*}
    \centering
    \includegraphics[width=\columnwidth]{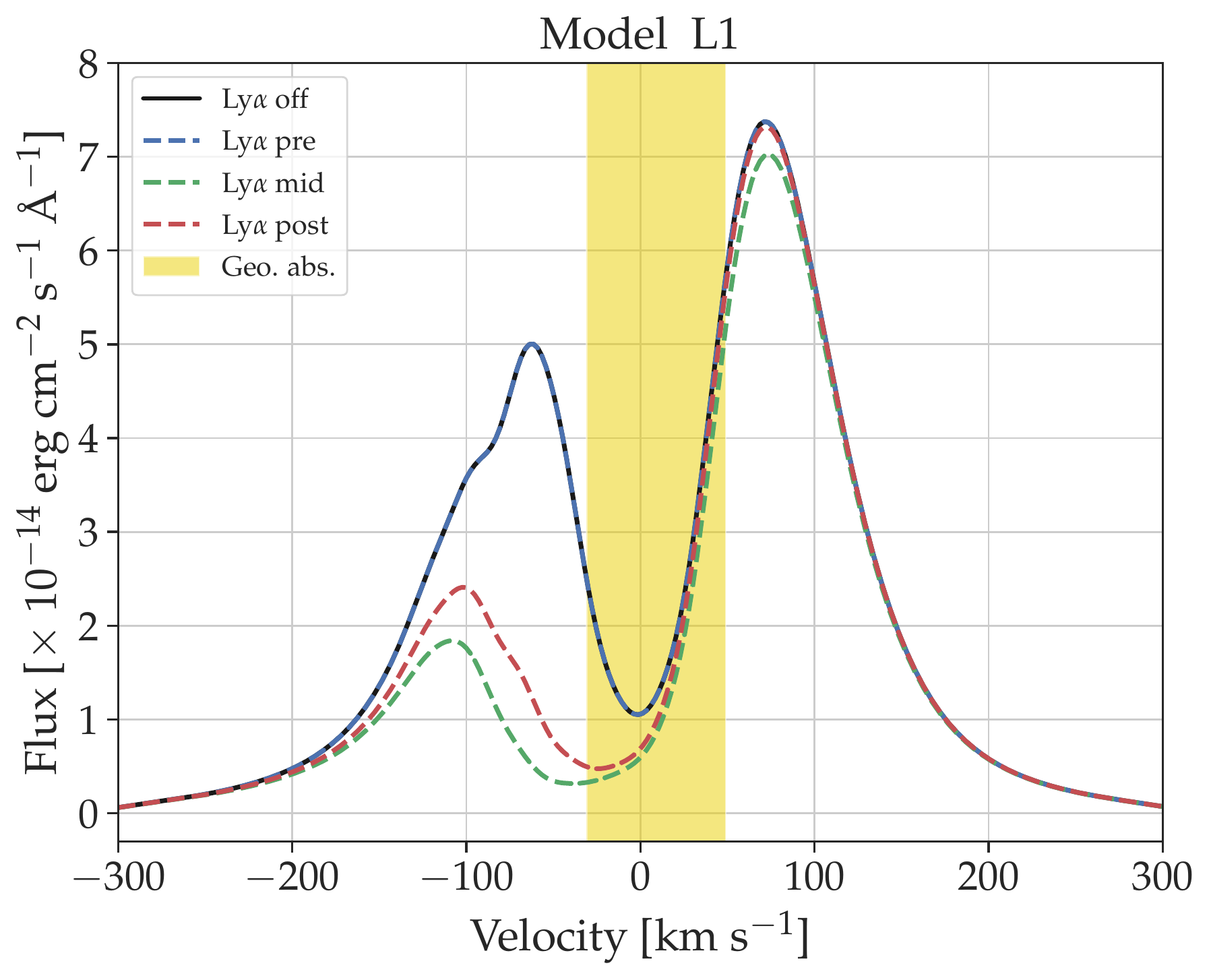}
    \includegraphics[width=\columnwidth]{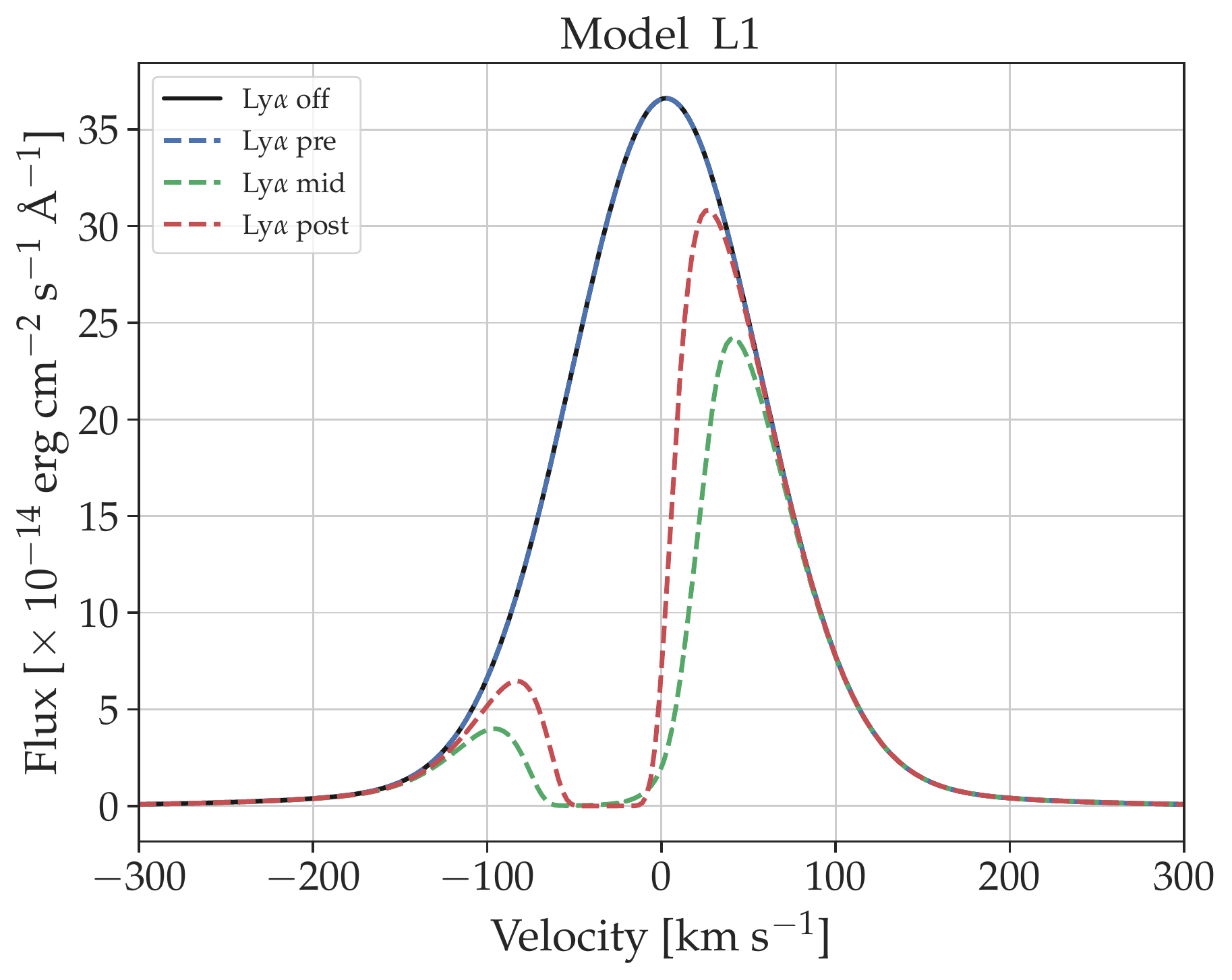}
    \caption{Lyman $\alpha$ flux as a function of the velocity measured from the centre of the line in the stellar reference frame. The different lines represents the 
flux out of transit (filled black), 2 hours before mid-transit (blue dashed), at mid-transit (green dashed) and 2 hours after mid-transit (red dashed). The yellow 
stripe represents the part of the line contaminated with the geocoronal emission and interstellar absorption. The right panel includes the effect of ISM attenuation 
along GJ 436's sight line and is convolved to match the STIS G140M resolution, and in the left panel no ISM attenuation and high spectral resolution is 
assumed.}
    \label{fig:lya_models}
\end{figure*}

The  strongest (and most robust) feature of the Ly$\alpha$ observations happen in the blue wing and, with this in mind, models L1 and M3 are the ones that 
better reproduce the observations.
To simulate an observation made with the STIS instrument, we convolved our Ly$\alpha$ profile (already attenuated by the ISM and absorbed by the neutrals) 
with the line spread function of the G140M grating. 
We show in Figure \ref{fig:lya_models} the Ly$\alpha$ line profile produced by model L1 at four different times: out of transit (black), 2 hr before mid-transit 
(blue), at mid-transit (green) and 2 hr after mid-transit (red) as shown in \cite{Ehrenreich2015}. We also show in Figure~\ref{fig:lya_models} what the in- and out-of-transit spectra would look like at high spectra resolution without any ISM attenuation or geocoronal contamination.

{}

\subsection{H$\alpha$ calculation}
We obtained the H$\alpha$ synthetic transit profile using the same approach than for the Ly$\alpha$ calculations. The optical depth computation is as explained in Section \ref{lyacalc}, except that now we use the H$\alpha$ line parameters shown in Table \ref{tab:lines}. To estimate the population of  neutral H atoms with electrons on the first excited level or $l=2$ (with $l$ denoting the energy level of the atom), we follow a similar approach to \cite{Christie2013} and we refer the reader to this paper for a more in-depth discussion. 
In summary, the physical mechanisms that are included in the calculations of the number density of atoms at $l=2$, n$_{l=2}$, are: the electron and proton collisional excitation and de-excitation, spontaneous radiative decay, photoexcitation and stimulated emission.  We do not include recombination or photoionisation from the 2nd level, and as calculated for HD189733b by \cite{Christie2013} and \cite{Huang2017} they becomes negligible compared to other dominant processes such as radiative decay and photoexcitation. 

Solving for the level 2 population is done at post-processing employing the {\it populate()} method within the subroutine {\it ch.ion()} in the ChiantiPy package\footnote{https://chianti-atomic.github.io/index.html} \citep{Dere2019}. The {\it populate()} method computes the population of the different levels within an atom as a function of the temperature and the electron density including an external radiation flux. 
For the H$\alpha$ calculation we take the total number density of level 2 which is the sum of the levels 2s ($\mathrm{^2S_{1/2}}$) and 2p ($\mathrm{^2P_{1/2}}$ and $\mathrm{^2P_{3/2}}$).
We take the stellar  Ly$\alpha$ flux as the external radiation field, by approximating it by a Planck function with temperature $T_{{\rm Ly}\alpha,\star}$. We use $T_{{\rm Ly}\alpha,\star} = $ 8150 K, which gives almost the same value on average than the observed Ly$\alpha$ flux. 
This temperature resembles the chromospheric temperature found in the model of  \cite{Peacock2019} for GJ 436 and is a larger than the 7000K assumed for the solar like star HD189733  in \cite{Christie2013} and more similar to the 8000K adopted in \cite{Huang2017}. As in \cite{Christie2013} we are also neglecting  the attenuation, difussion and re-emission of the photons as they travel through the planetary material, otherwise, a full radiative transfer computation as the one done by \cite{Huang2017} should be included, and this is outside the scope of this work. The dilution factor for the external radiation field is $W = 0.5 [1 - (1-R_\star^2/a_p^2)^{1/2}]$, with $a_p/R_\star = 12.5$. 

To compare with the observational results from \citet{Cauley2017}, we compute the  equivalent width of the H$\alpha$ line using
$$W_{H_\alpha}=\sum_{v=-200}^{200}(1-I_v/I_\star)\Delta\lambda_v, \label{eqwidth}$$ 
where $\Delta\lambda_v$ is the wavelength difference at velocity $v$. We note that the integration is done in the velocity range $\pm 200$ km s$^{-1}$ as in 
\citet{Cauley2017}.  

\begin{figure}
    \centering
    \includegraphics[width=\columnwidth]{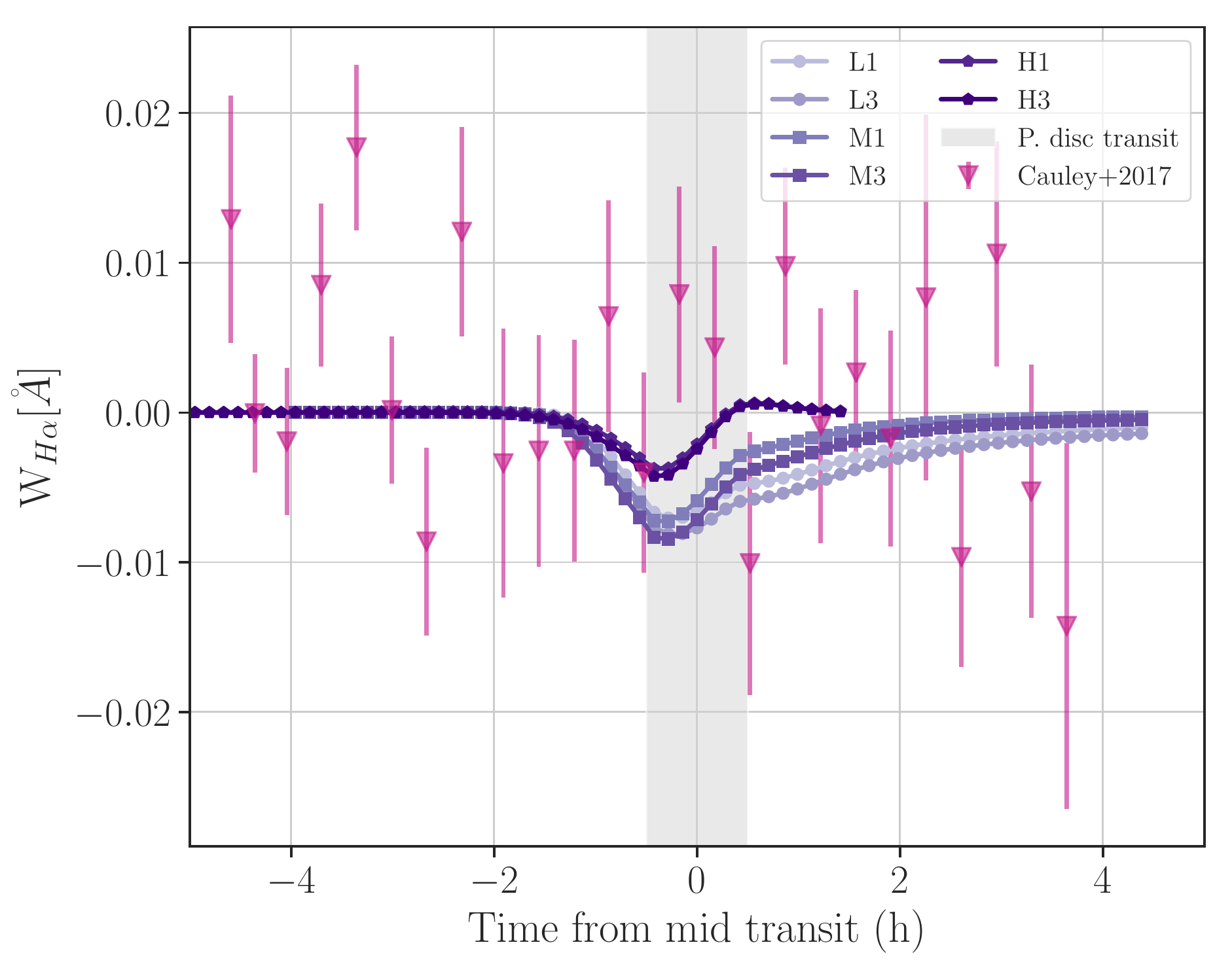}
    \caption{Equivalent width of H$\alpha$~as a function of time from mid-transit for all our models. Pink triangles in show the observations from 
\protect\citet{Cauley2017}. The grey band represents the duration of the optical transit.}
    \label{fig:ha_models}
\end{figure}

The equivalent widths we compute for each of our simulations are shown in Figure \ref{fig:ha_models} as a function of time from mid-transit, with the grey band 
marking the duration of the optical transit. The observations from \citet{Cauley2017} are the pink data points, revealing no H$\alpha$ absorption. Figure 
\ref{fig:ha_models} shows that the absorption produced by our models is, in all cases, within the error and the dispersion found in the observation. Hence, we 
conclude that all of our models agree with the non detection of H$\alpha$ absorption for this system.   

\citet{Cauley2017} estimated that the ratio N$_{l=2}/\mathrm{N}_{l=1}$, where N$_{l=2}$ is the column density of hydrogen atoms in the first excited level and 
N$_{l=1}$ the column density of H atoms in the ground state, should yield a value $ < 10^{-3}$ in order for the absorption to be detected by the observations. 
Our computed values for the exited-to-ground-state ratios are around $4\times 10^{-7}$, well below the upper limit estimation from \citet{Cauley2017}. 

{}

\section{Discussion} \label{summ}
In this work, we simulated the 3D interaction between the escaping atmosphere of GJ 436b and the wind of its host star. We explored different values of stellar 
wind strengths (given by base temperatures of 1 and 3 MK) and EUV luminosities ($[0.8, 1.6, 4]\times 10^{27}$ erg s$^{-1}$). 
From our set of 3D simulations, we constructed synthetic observations of the transit of GJ 436b in two hydrogen lines: H$\alpha$ and Ly$\alpha$. 
In general, the interaction between the stellar and planetary winds resulted in a comet-like tail behind the planet and a shock ahead of the planet. The neutral 
material that leaves the planet's atmosphere is then distributed within this region and is responsible for the absorption seen in the Ly$\alpha$ line. Because this 
material is asymmetrically distributed around the planet, the Ly$\alpha$ transit lightcurve is asymmetric, with different shapes depending on the EUV luminosity 
and stellar wind strength. 

Overall, the Ly$\alpha$ absorption correlates more strongly with the stellar EUV flux in the blue wing -- we found a larger absorption depth for lower EUV 
luminosities. In the red wing of the Ly$\alpha$ line, the stellar EUV flux does not have a great impact as most of the models give the same absorption depth. 
The absorption depth in the blue wing of the Ly$\alpha$ line is also dependent on the stellar wind strength. The stellar wind shapes the distribution of the 
material that leaves the planet and so, the absorption depth is larger  for a stronger wind (considering the same L$_{\rm EUV}$). The same behaviour can be  
seen in the red wing.

The maximum depth and the duration of the absorption lightcurve in the blue wing of the line \citep{Lavie2017} is well reproduced by models 
L1 and M3. Model L1 has a low EUV luminosity and a stellar wind velocity of 254 km s$^{-1}$ at the planet position, while model M3 represents a more moderate L$_{\rm EUV}$ value and a stronger stellar wind with a velocity of 465 km s$^{-1}$ at the orbital distance.

The red wing absorption found after mid-transit is not properly reproduced by any of our models, as most of them produce rather symmetric lightcurves around mid-transit. The impossibility of fitting both red and blue absorption is also presented in \cite{Khodachenko2019}. 

The set of models with the larger L$_{\rm EUV}$ showed a different lightcurve in the blue wing. For these models a stream of material moving towards the star develops. Model H3 shows an early absorption in the blue wing of the line with a shallower depth and model H1 shows an early absorption in the red wing. In both cases, it is the material ahead of the planet the one producing the early ingress (also seen in \citet{Bourrier2015}).

However, these models cannot simultaneously reproduce the long-lived absorption seen after mid-transit in the blue wing. 
One possibility is that the early absorption might have been produced in a time variable event such as accretion of material towards the star occurring sporadically. 

Contrary to the huge absorption observed in  Ly$\alpha$, GJ 436b does not show any significant absorption in the H$\alpha$ line. To compare with the 
observations, we computed the synthetic H$\alpha$ absorption from our models. We found that all our models give a non detection of absorption during transit in 
agreement with the observations presented in \citet{Cauley2017}.

\subsection{Effects of charge exchange and radiation pressure in our simulations}\label{discussion_radp_cexch}
The high-velocity component the Ly$\alpha$ line profile at high velocities (Fig. \ref{fig:lya_models}) are due to radiation pressure accelerating neutral hydrogen atoms and the stellar wind ram pressure that drags and accelerates planetary particles. One additional process that can contribute to this high-velocity component is charge exchange process, which we neglected in our models. We discuss some of these process below. 

By running an extra set of simulations we investigated the role of radiation pressure in the models with the largest neutral density fraction at the planetary wind boundary (L1 and L3). A detailed analysis of these additional test simulations are seen in Appendix \ref{appendixA1}. By neglecting the radiation pressure, the effective gravity of the star is larger, which makes the planetary wind expand towards the star. The stellar wind provides the ram pressure necessary to stop this expansion. Thus, in the case of a weaker stellar wind (L1), this expansion (and consequent bow shock formation) is stopped further away from the planet than in the case of stronger wind (L3). For case L3, actually, radiation pressure affects very little the dynamics of the interaction, and thus the Ly$\alpha$ transit line profile is barely affected (see Figure \ref{fig:line_profile_nrp}). For case L1, on the other hand, radiation pressure  plays a larger role. This occurs because in this simulation the planetary wind neutral density is large and the stellar wind is not too strong.

Several authors have shown that radiation pressure does not play an important role in the case of HD 209458b and GJ 436b \citep[among others]{Cherenkov2018, Esquivel2019, Khodachenko2019, Debrecht2020}. However, we have found  here that this might not be always the case as it depends on the stellar wind and planetary wind parameters together with the flux in Ly$\alpha$.

One limitation of our treatment is that our models do not include self-shielding of neutrals from the Ly$\alpha$ photons. With self-shielding, the stellar Ly$\alpha$ photon that is absorbed in a region of the planetary atmosphere can no longer contribute and thus  radiation pressure weakens as one goes deeper into the planetary atmosphere. As a result, neglecting self shielding leads to an overestimation of the radiation pressure in regions with a large amount of neutrals.
In Appendix \ref{appendixA2}, we estimate the effect of neglecting self-shielding would have in our results. We  found that this affects mostly the regions near the planet, where the planetary wind has smaller radial velocities. These regions only contribute to the Ly$\alpha$ transit near line center ($\pm 30$ km/s), which are excluded from the Ly$\alpha$ analysis due to geocoronal emission. Although they could be important for the H$\alpha$ analysis, the lack of absorption in this line also makes self-shielding  not important. This is similar to what was concluded in the work of \cite{Esquivel2019} for HD 209458 and by \cite{Khodachenko2019} in the case of GJ 436.

Our models do not consider the processes of charge exchange. Inclusion of this phenomenon is well addressed by \cite{Khodachenko2019} in a multi-fluid simulation for GJ 436b. In their work, the authors attributed the blue-shifted absorption in synthetic Ly$\alpha$ transits mainly due to charge exchange. Even though the effect of charge exchange has been modelled (approximately) in the past with single fluid simulations \citep{Tremblin2013,Esquivel2019}, we have chosen not to follow this path, because in order to have a realistic effect a multi-fluid treatment is necessary.

However, since charge exchange is produced near the shock between the stellar and planetary winds, adding this phenomenon to our models would have shifted the maximum absorption in the blue wing towards a time after mid-transit  (as seen in figure 10 of \citealt{Khodachenko2019}). The total amount of neutral material would remain the same, but the newly created neutral atoms would have the velocity of the stellar wind. 

\section{conclusions}\label{conc}
We have created a set of 3D models for the wind interaction in the exoplanetary system GJ436 varying the strength of the stellar wind and the stellar L$_\mathrm{EUV}$. Based on spectroscopic transit 
observations we explored different scenarios for the winds in the GJ 436 system.
If we constrain our models based only on the Ly$\alpha$ observations of GJ 436b, we found that there is an ambiguity in the number of models that can 
reproduce these observations. In this case, models M3 and L1 can fit most of the observational points specially in the blue wing. However, we cannot 
disentangle which model is the best one. This degeneracy could be broken by using information from the synthetic H$\alpha$ transits. But, in the case of GJ 
436b, all our models agree with the non-detection of H$\alpha$ absorption during the planetary transit.

Simultaneously fitting multiple observational diagnostics can help us constrain models to better derive physical parameters of planetary systems.
In this work, we aimed to fit two spectral lines (H$\alpha$ and Ly$\alpha$) observations with a unique model but, more than one model was able to reproduce 
the H$\alpha$ observation making it hard to select our `best' model. Based on the Ly$\alpha$ results,  we conclude that the conditions of the stellar wind at the 
planet position are those determined by the values found in models L1 and M3: velocity $\sim [250-460]$ km s$^{-1}$, temperature of $\sim [4-3]\times10^5$ K and a stellar mass-loss rate of $2\times 10^{-15}$ M$_\odot$ yr$^{-1}$. The stellar EUV luminosity for this model is set at $[0.8-1.6]\times10^{27}$ erg s$^{-1}$, 
which produces a planetary mass-loss rate of $\sim [6-10]\times10^9$ g s$^{-1}$.

\section*{Data availability}
The data underlying this article will be shared on reasonable request to the corresponding author.

\section*{Acknowledgements}
We thank the referee for the comments. CVD acknowledge the funding from the Irish Research Council through the postdoctoral fellowship (Project ID: GOIPD/2018/659) and postdoctoral fellowship from SECYT-UNC 2020.  AAV and GH have received 
funding from the European Research Council (ERC) under the European Union's Horizon 2020 research and innovation programme (grant agreement No 
817540, ASTROFLOW). AE acknowledges support from the the DGAPA-PASPA (UNAM) program. The authors wish to acknowledge the SFI/HEA Irish Centre 
for High-End Computing (ICHEC) for the provision of computational facilities and support. 


\bibliographystyle{mnras}
\bibliography{villarreal_gj436_final} 


\appendix
\section{The effect of radiation pressure in our models} \label{appendix}
\subsection{Comparison between models with and without radiation pressure}\label{appendixA1}
Here we investigate the effects that our implementation of the radiation pressure force has on our models. For this purpose we run a new set of simulation for models L1 and L3 without radiation pressure. In our setup this means we set $\beta=0$ in Equation \ref{eq:grav_eff}. We choose these models as they are the ones with the highest neutral planetary density at the planet boundary condition, where we expect  the effects of radiation pressure to be more noticeable.  

\begin{figure}
\begin{center}
\includegraphics[width=\columnwidth]{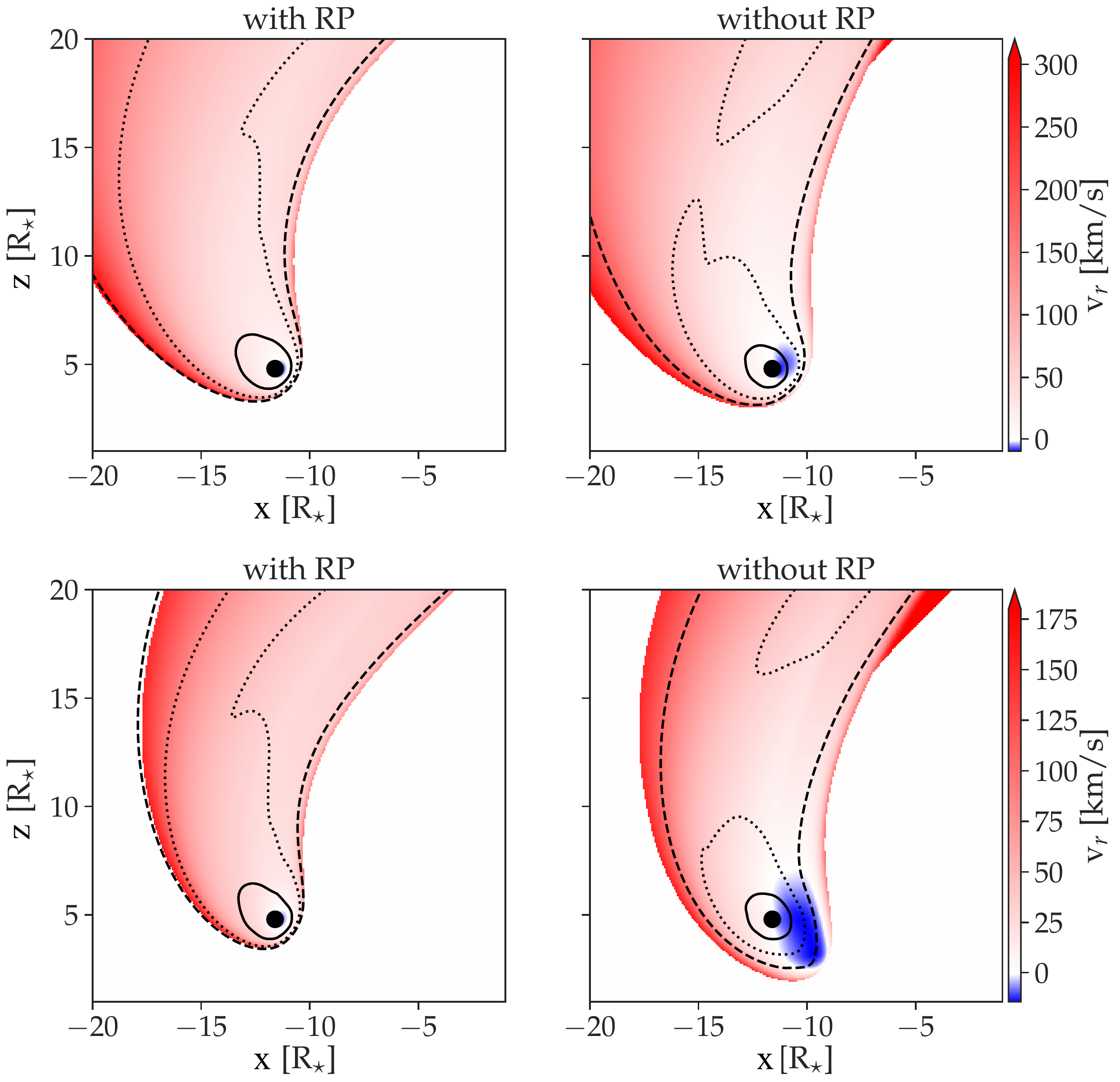}
\caption{Radial velocity distribution in the orbital plane for model L1 (top) and model L3 (bottom) with (left) and without (right) radiation pressure. Velocities are measured from the stellar reference frame. Contours of ionisation fraction are show in black: solid contour for 0.6, dashed for 0.8 and dotted for 0.9.  Black filled circle shows the position of our boundary  for the planetary wind (5 R$_p$).}
\label{fig:rad_profiles_radp}
\end{center}
\end{figure}

Figure \ref{fig:rad_profiles_radp} shows the radial velocity distribution in the orbital plane for model L1 with (left panels) and without (right panels) radiation pressure. Negative velocities, i.e. towards the star, cover a larger area for model without radiation pressure. This is a direct consequence of the larger effective stellar gravity now felt by the neutrals. 
Material being accelerated towards the star now shows a shock further from the planet in comparison with the model that includes radiation pressure. 
The lack of radiation pressure in model L3 (bottom panels of Fig. \ref{fig:rad_profiles_radp})  has a less dramatic effect, with the amount of material with negative velocities much closer and within a smaller region around the planet. 
 
Because there is no much difference in the ram pressure of the planetary wind compared to the case with radiation pressure, the position of the shock does not change too much between these two models.

\begin{figure}
\begin{center}
\includegraphics[width=0.7\columnwidth]{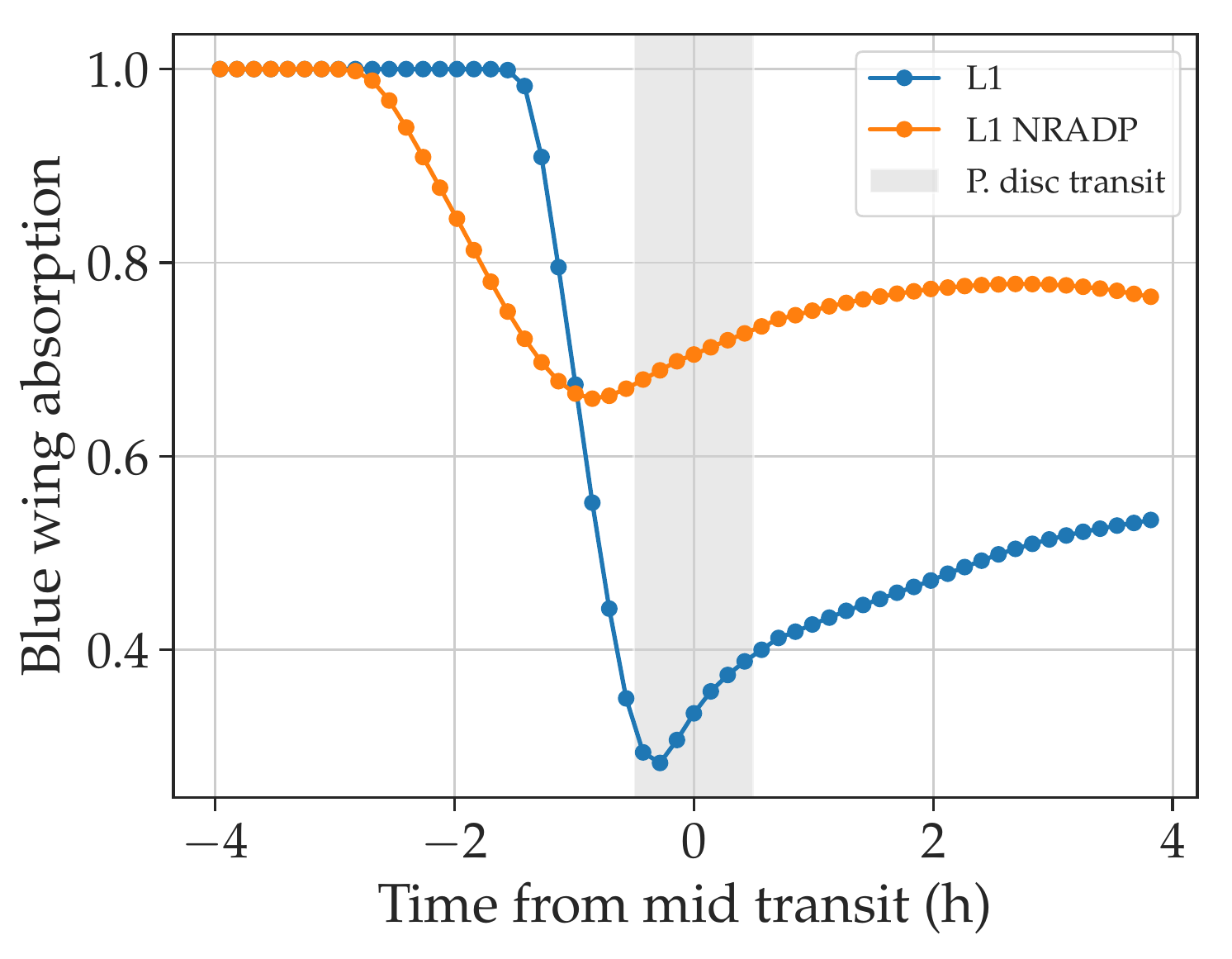}
\includegraphics[width=0.7\columnwidth]{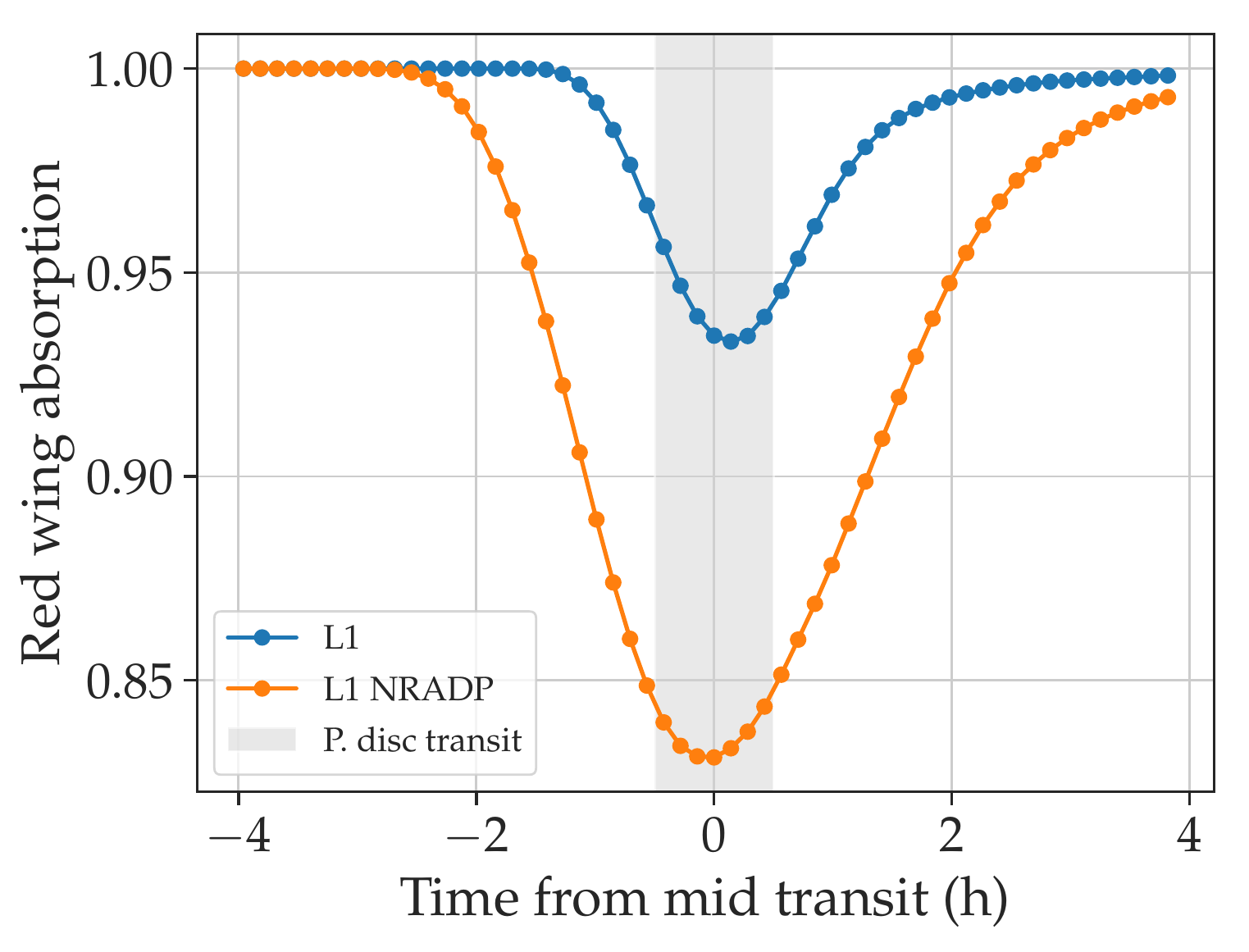}
\caption{Blue ([-120, -40] km s$^{-1}$) and red ([30,100] km s$^{-1}$) wing absorption in Ly$\alpha$ as a function of time from mid-transit for model L1 with (light blue lines) and without (orange lines) radiation pressure. The grey band represents the duration of the optical transit. }
\label{fig:lightcurve_nrp}
\end{center}
\end{figure}

\begin{figure}
\begin{center}
\includegraphics[width=0.7\columnwidth]{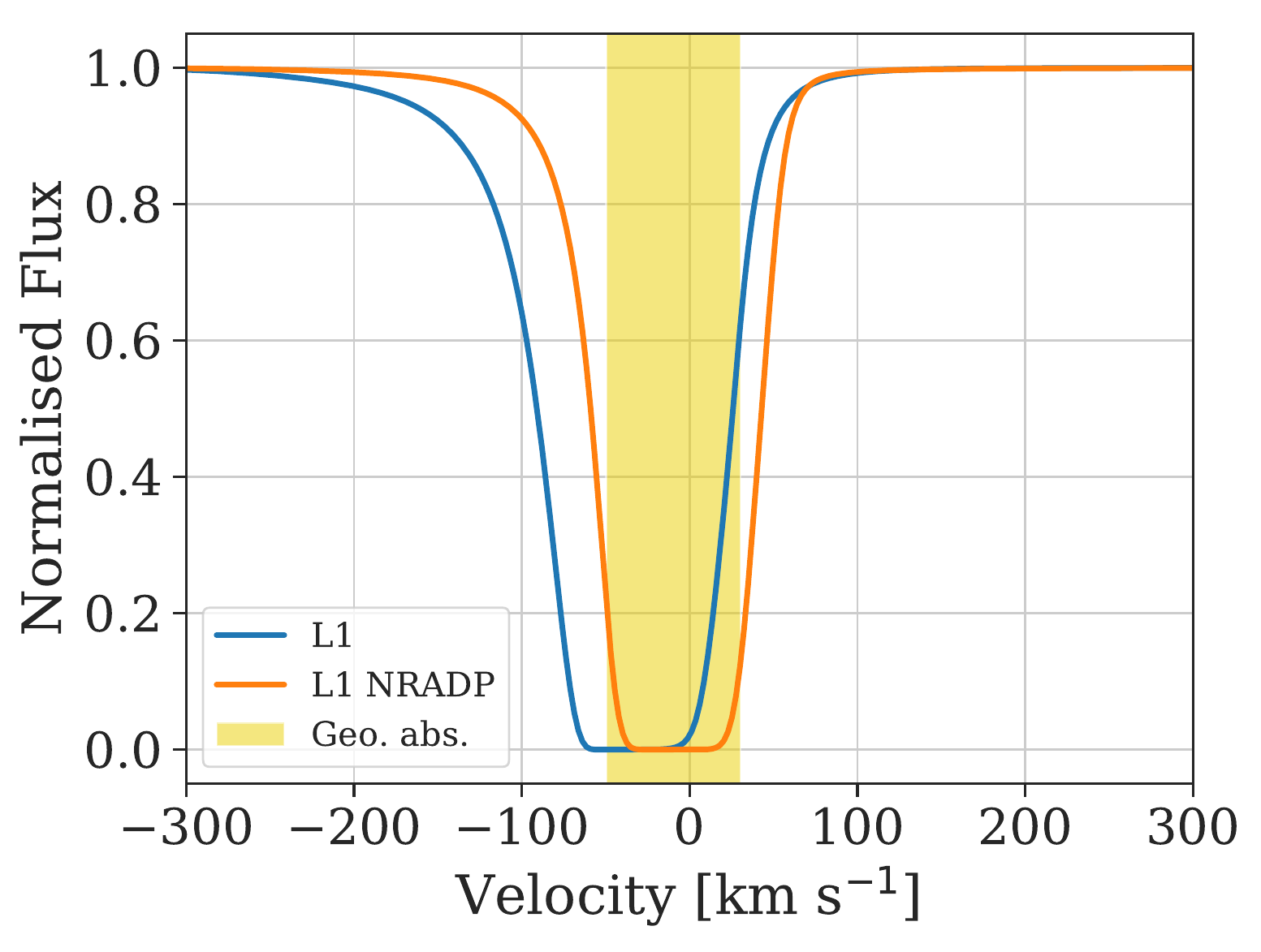}
\includegraphics[width=0.7\columnwidth]{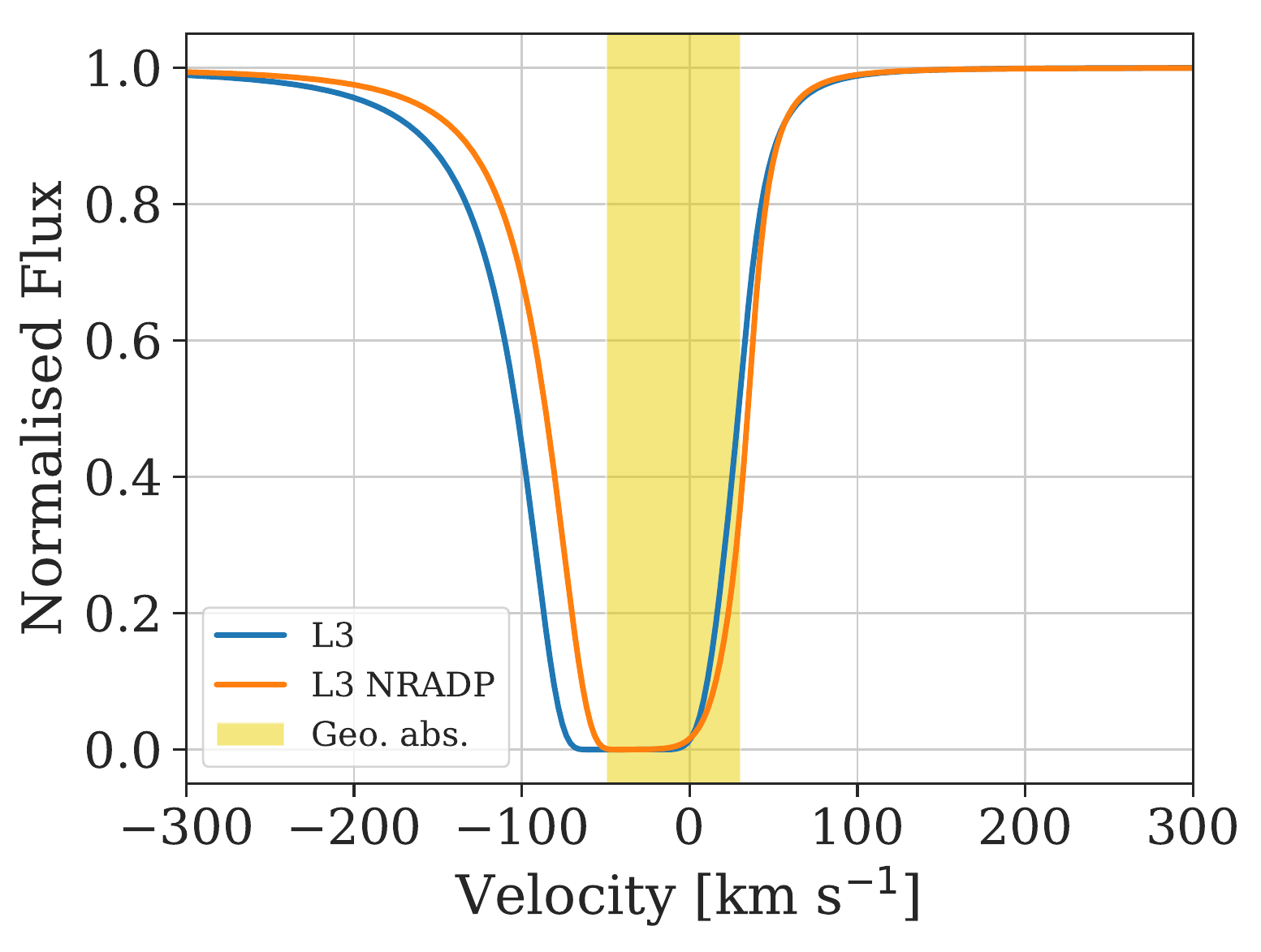}
\caption{Normalised flux in the Ly$\alpha$ line as a function of velocity at mid-transit for model L1 (top panel) and L3 (bottom panel) with (light blue) and without (orange) radiation pressure. The signal is in the heliocentric reference frame. The yellow band shows the part of the line contaminated with the geocoronal emission.}
\label{fig:line_profile_nrp}
\end{center}
\end{figure}

In the case were radiation pressure makes a more significant difference in the velocity distribution of neutrals (model L1), we also investigate the resulting Lyman $\alpha$ profile during transit. Figure \ref{fig:lightcurve_nrp} shows the absorption in both wings of the Ly$\alpha$ line as a function of time from mid-transit for the cases with (light blue lines) and without (orange lines) radiation pressure. Since part of the material that moves towards the star remains neutral (see top right panel of Figure \ref{fig:rad_profiles_radp}), the absorption in the red wing is larger compared to the case with radiation pressure. On the contrary,  in the blue wing, the absorption has a smaller depth but starts a couple of hours earlier than in the case with radiation pressure. Also, when computing the line profile at mid transit in Figure \ref{fig:line_profile_nrp}, we can see that without radiation pressure, the profile is red shifted due to the neutral material moving towards the star, being more pronounce for model L1 (left panel in Figure \ref{fig:line_profile_nrp}) than in model L3 (right panel Figure \ref{fig:line_profile_nrp}).

From our test models, we have seen that radiation pressure acts on the densest regions of neutral material, and when it is not present, the planetary wind more easily expands towards the star. This behaviour is suppressed when the stellar wind ram pressure is strong enough to balance the planetary wind ram pressure at a closer distance to the planet, like in the case of model L3. 

In model L1,  the line profile at mid-transit shows a larger absorption at high negatives velocities when radiation pressure is included but there is not too much difference in model L3, when the stellar wind is stronger. This implies that in this last model, the stellar wind is mainly the responsible for producing the blue-shifted high velocity neutrals while in model L1 radiation pressure force is also responsible.

\subsection{Self-shielding} \label{appendixA2}
As we mentioned in section \ref{grp}, one caveat of our radiation pressure implementation is the lack of self-shielding. To study the influence of self-shielding in our models we have followed the same procedure as in \cite{Esquivel2019} and compute the attenuation of the Lyman $\alpha$ flux due to the column density of neutral material in the path of the Ly$\alpha$ photon for different velocity bins. Even though this study is done as post-processing, since it will require a larger computational time to include self-shielding at real time, it gives us an idea of how much the radiation pressure within a cell  changes if we allow the absorption (shielding) by neutrals. 

The optical depth for this case is calculated by integrating the neutral column density from the star to the position of a given cell in velocity intervals, $\Delta v =15$ km s$^{-1}$: 
\begin{equation}
\tau(v_{r,l}) = \frac{\sigma_{0} \lambda_0}{\Delta v} \int^l_0 n_\mathrm{H}(v_r,l') dl'
\end{equation}
where $\sigma_{0}$ is the absorption cross-section at the threshold wavelength and $ \lambda_0$ is the Ly$\alpha$ central wavelength. Here, we are assuming that the line profile is a Delta function centred at the velocity $v_r \pm \Delta v/2$. 

\begin{figure}
\begin{center}
\includegraphics[width=\columnwidth]{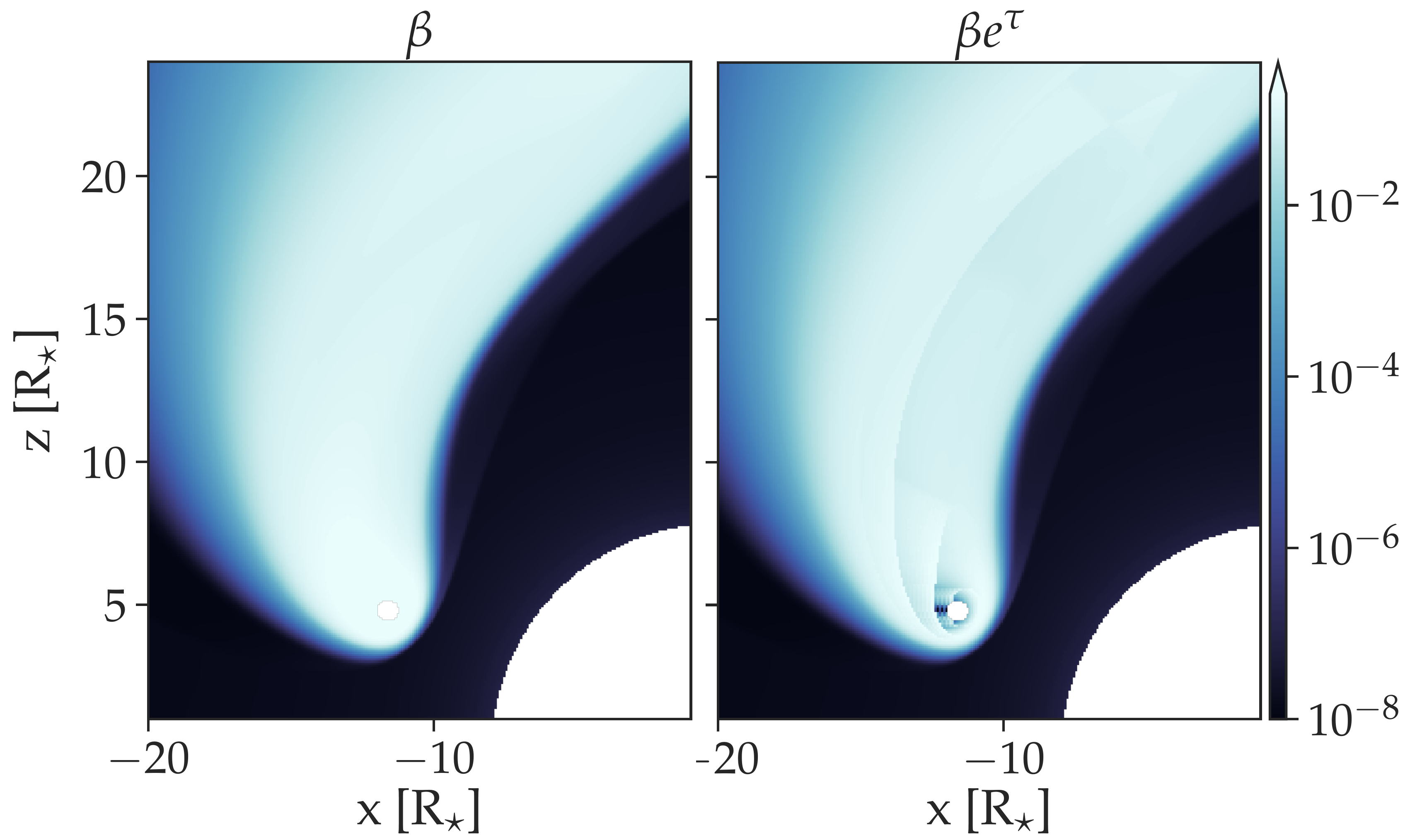}

\caption{$\beta$ value distribution in the orbital plane of model L1 without (left) and with (right) self-shielding. Withe circles are the boundary condition regions for the stellar wind and planetary wind. }
\label{fig:betatau}
\end{center}
\end{figure}

Figure \ref{fig:betatau} shows the $\beta$ value for model L1 as is used in our simulations (left panel) and the resulting value when including self-shielding in post-processing (right panel). We can see that $\beta$ is reduced near the planet and in the inner regions of the cometary tail where the larger amount of neutrals are found. 
However, the largest attenuation is seen in a region very close to the planet where radial velocities are less than 30 km s$^{-1}$ (this conclusion was also found by \citealt{Khodachenko2019}). 
This velocity range is excluded from our Ly$\alpha$ transit calculations, as they belong to the Ly$\alpha$ region that is contaminated with the ISM absorption and the geocoronal emission. 
Thus, in our models, including self-shielding would only affect the regions of the Ly$\alpha$ line that is not observable. However, other systems might not behave as GJ 436 and thus to accurately see the effect of self-shielding, it is preferable to directly include it in the simulations.

Additionally, we can say that for model L1, the real distribution of the neutral material will be something  between the cases with and without radiation pressure. Combined to the conclusions presented in Appendix \ref{appendixA1}, when the neutral density is small and/or the stellar wind is stronger, radiation pressure will be less important and neglecting self-shielding will have an even lesser effect in our results.

\bsp	
\label{lastpage}
\end{document}